\documentclass[prd,aps,twocolumn,floats,floatfix,nofootinbib]{revtex4-1}

\usepackage{amssymb,bm}
\usepackage{graphicx}
\usepackage{dcolumn}
\usepackage{bm}
\usepackage{graphics}
\usepackage{slashed}
\usepackage{enumerate}
\usepackage{amsmath}
\usepackage{hyperref}
\usepackage{url}
\usepackage[usenames,dvipsnames,svgnames,table]{xcolor}
\usepackage{color}
\usepackage{xcolor}
\usepackage{grffile}
\usepackage{booktabs}
\usepackage[toc,page]{appendix}
\usepackage{float}
\usepackage{soul}
\usepackage{subfigure}
\usepackage{multirow}
\usepackage{booktabs}

\hypersetup{
	colorlinks=true,
	linkcolor=red,
	linkbordercolor=black,
	citecolor=blue,
	filecolor=magenta,      
	urlcolor=magenta,
	pdfpagemode=FullScreen,
}

\newcommand{\tmmathbf}[1]{\ensuremath{\boldsymbol{#1}}}

\newcommand{\Mycite}[1]{\citeauthor{#1}~(\citeyear{#1})}
\newcommand{\Mycitep}[1]{\citeauthor{#1}~\citeyear{#1}}

			{\end{center}\end{minipage}\end{flushleft}} 
		

\begin{document}
\title{Global high-order numerical schemes for the time evolution of the general relativistic radiation magneto-hydrodynamics equations}
\author{ M.~R.~Izquierdo$^{1,2}$, L.~Pareschi$^{3}$, B.~Mi\~nano$^{2}$, J. Mass\'o$^{1,2}$ and C. Palenzuela$^{1,2}$}
	
	\affiliation{${^1}$Departament  de  F\'{\i}sica,  Universitat  de  les  Illes  Balears,  Palma  de  Mallorca, E-07122,  Spain}
	\affiliation{$^2$Institute of Applied Computing \& Community Code (IAC3),  Universitat  de  les  Illes  Balears,  Palma  de  Mallorca, E-07122,  Spain}
	\affiliation{$^3$Department of Mathematics and Computer Science, University of Ferrara, Ferrara, I-44121, Italy}
%
\begin{abstract}
	Modeling correctly the transport of neutrinos is crucial in some astrophysical scenarios such as  core-collapse supernovae and binary neutron star mergers. In this paper, we focus on the truncated-moment formalism, considering only the first two moments (M1 scheme) within the \emph{grey} approximation, which reduces Boltzmann seven-dimensional equation to a system of $3+1$ equations closely resembling the hydrodynamic ones. Solving the M1 scheme is still mathematically challenging, since it is necessary to model the radiation-matter interaction in regimes where the evolution equations become stiff and behave as an advection-diffusion problem. Here, we present different global, high-order time integration schemes based on Implicit-Explicit Runge-Kutta (IMEX) methods designed to overcome the time-step restriction caused by such behavior while allowing us to use the explicit RK commonly employed for the MHD and Einstein equations. Finally, we analyze their performance in several numerical tests.
\end{abstract} 
%
\maketitle
%
%
\normalsize
%
\section{Introduction}
The concurrent observation of gravitational waves (GW) and a wide range of electromagnetic (EM) signals in the event GW170817/GRB170817A/AT2017gfo started an exciting era of multimessenger astronomy (\Mycitep{2041-8205-848-2-L12}; \Mycitep{PhysRevLett.119.161101}). These combined observations provided exquisite information on the source, a binary neutron star (BNS) merger,  allowing us to tackle some fundamental questions in physics such as the behavior of very high-dense matter present at the neutron star (NS) cores. The complexity on modeling all the physical interactions involved constraints our ability to realistically simulate these binary mergers and compare them with current observations. Decoding the physics imprinted on these signals requires solving, at least approximately, the general relativistic radiation magneto-hydrodynamic  equations:  Einstein equations to describe the strong-gravity, relativistic magneto-hydrodynamics (MHD) for the magnetized fluid, and Boltzmann's equation for the production and transport of neutrinos. Finding accurate solutions of these  equations in realistic astrophysical scenarios such as BNS mergers is an extremely difficult task which can only be approached through numerical simulations. 

Although many of these physical ingredients have been incorporated routinely in BNS simulations, an accurate treatment of the neutrinos has remained yet elusive. The effects of neutrinos are crucial to model the kilonova, a strong EM radiation powered by the radioactive decay of heavy r-process nuclei in the material ejected from the stars during and after the merger. In particular, the neutrinos largely determine the amount and composition of the material ejected long after the merger (i.e., secular ejecta), which is responsible for part of the kilonova emission (see, e.g., \Mycitep{sekiguchi15}; \Mycitep{foucart16a}; \Mycitep{radice16}; \Mycitep{sekiguchi16}; \Mycitep{perego17}). 

Neutrinos can be described by a distribution function  obeying the Boltzmann equation, which accounts for the transport and interactions of the neutrinos. In the most general case, this equation is characterized by the spatial and time coordinates, the neutrino energies, and their direction of propagation (two angles), which makes the problem 7-dimensional. In astrophysical systems such as core-collapse supernovae and remnants of BNS mergers the neutrinos commonly propagate through dense regions, where the radiation diffuses, and then across low-density regions, where it streams freely. These two different behaviors are reflected in the neutrino transport equation, which becomes parabolic in the diffusion limit and hyperbolic in the free streaming one (\Mycitep{pomraning83}; \Mycitep{mihalas84}). Obviously, in order to model the  neutrino dynamics accurately, we need not only to capture correctly these two limiting regimes, but also the transition region between them. This is an extremely challenging computational 7-dimensional problem which requires to rely on approximations and simplifications to make it manageable. Although there have been many others, mainly three approaches have been commonly employed to model efficiently the neutrino transport, with different accuracy results: the leakage scheme (\Mycitep{ruffert96}; \Mycitep{rosswog03}), the (truncated) moment formalism (\Mycitep{thorne81}; \Mycitep{shi11}) and Monte Carlo algorithms (\Mycitep{miller19a}; \Mycitep{Foucart_2020}). For a detailed review and comparison of these methods we recommend the recent work by \Mycite{foucart_review}.

The simplest and most inexpensive approach is the neutrino leakage scheme (see, e.g., \Mycitep{Sekiguchi_2010}; \Mycitep{deaton13}; \Mycitep{illinois_leakage22}; \Mycitep{leakage22}). This approximation seeks to account for changes in the electron fraction and in the energy losses due to the emission of neutrinos. The leakage scheme replaces the transport equations with an estimation of the cooling rate based on the optical depth, which requires a minimization of the path integral of the opacity in all possible directions. Although this is an expensive non-local calculation, the shortest distance between a point and a surface-level can also be calculated efficiently by solving instead the eikonal equation (\Mycitep{Neilsen:2014hha}; \Mycitep{leakage22}). Although many groups (see refs. in \Mycitep{foucart_review}) have employed this approach in the last decades, the lack of neutrino re-absorption might produce in some scenarios only crude estimations of the amount and composition of the ejecta.

The most accurate but still manageable option consists on solving the 7-dimensional Boltzmann equation by using Monte Carlo methods (see, e.g., \Mycitep{Roth_2022}; \Mycitep{kawaguchi22}; \Mycitep{fou22}), which are particularly efficient for high-dimensional problems. In this approach,
the distribution function of neutrinos is discretized using super-particles, each representing a large number of neutrinos. The emission and absorption of these  super-particles are treated probabilistically, while they propagate just along geodesics. This approach is not only the most accurate of those discussed here, but it is also the only one that converges to the true continuum solution in the limit of infinite number of particles. However, its cost is still considerably larger than the other approaches, especially when dealing with optically thick regions, making them very expensive and not yet commonly employed in long simulations.

An intermediate approach, both in accuracy and in computational cost, relies on evolving the moments of the distribution function of the neutrinos (see, e.g., \Mycitep{fou16b}; \Mycitep{weih20}; \Mycitep{rad2022}; \Mycitep{zappa22}). Furthermore, two simplifying approximations are commonly applied: (i) the moments are truncated, such that an algebraic closure needs to be provided in order to compute the higher moments, and (ii) evolution equations for the energy-integrated moments are obtained, averaging over the neutrino energies. In this way, the 7-dimensional Boltzmann equation is reduced into a $3+1$ system that resembles the hydrodynamic equations. Unfortunately, even this simplified scenario still presents numerical challenges. In particular, the truncated moment equations contain potentially stiff source terms, arising from the interaction of neutrinos with the matter, such that in optically thick regions the equations change from hyperbolic to parabolic type.

One way to overcome this issue is by using a spatial discretization scheme, based on the flux-splitting approach \Mycite{leveque}, that remains consistent also in the diffusion limit, while using some type of implicit methods for the time integration \Mycite{rad2022}. Actually, many recent works implementing these equations involve semi-implicit or Implicit-Explicit (IMEX) Runge-Kutta (RK) time integrators (\Mycitep{ascher97}; \Mycitep{pareschi00}; \Mycitep{kennedy03}; \Mycitep{pareschi05}). These methods can be very efficient for this type of equations, since only the potentially stiff terms in the neutrino transport equations need to be solved implicitly, while all the others can be treated explicitly. The complexity of the stiff terms and the moderate effect of the neutrinos in the General Relativity (GR) and MHD equations have lead to simple choices of the implicit time integrators to evolve the neutrinos, trying to minimize the number of evaluations of the stiff terms at the cost of getting only first-order accurate schemes.
This produces a mismatch with the Einstein and MHD equations, which are usually integrated by employing at least third order explicit Runge-Kutta methods. Therefore, there is a dichotomy of time integrators (i.e., a low-order semi-implicit for neutrinos and explicit high-order RK for the spacetime and fluid) that is producing a degradation on the accuracy of the full solution and/or an increase in the computational cost due to the time interpolations required to pass information between the neutrinos and the other fields in intermediate steps during the time integration. These problems only get worse in simulations using Adaptive Mesh Refinement (AMR), since the solution has to be interpolated both in space and time at the boundaries of the refinement levels.

The truncated moment formalism and the spatial discretization methods presented here are analogous to others in the literature (see, e.g., \Mycitep{fou16b}; \Mycitep{weih20}; \Mycitep{rad2022}). However, we present for the first time a global (i.e., it can be employed for all the equations involved in the problem, not only the stiff ones), high-order (i.e., fourth order accuracy for the explicit terms and up to third order for the implicit ones), time integrator that makes no trade-offs between performance and accuracy. The main difference lies in the design of an IMEX Runge-Kutta that satisfies a stringent list of requirements, such as achieving high-order accuracy and allowing the evaluation of the equations at the same intermediate times to avoid interpolation.  We would like to stress that this general framework can be applied not only to combine the spacetime and fluid evolution equations with the truncated moment formalism, but with any other system of equations with potentially stiff terms.

We have implemented these Runge-Kutta time integrators in the high-performance, open-source code \texttt{MHDuet} \Mycite{mhduet}. This code is generated automatically by the software \texttt{SIMFLOWNY} (\Mycitep{sim1}; \Mycitep{sim2}; \Mycitep{sim3}) and runs under the mature infrastructure \texttt{SAMRAI} (\Mycitep{samrai1}; \Mycitep{samrai2}), which has been shown to achieve exascale for simple problems. Currently, \texttt{MHDuet} solves the CCZ4 formalism of Einstein equations coupled to the MHD equations. The fluid pressure depends on the other thermodynamical quantities via a microphysical, tabulated equation of state. In order to accurately resolve the magnetohydrodynamical instabilities arising during BNS mergers, it also includes a technique to take into account the effect of the small unresolved scales into the large resolved ones, namely large-eddy simulations with the  gradient sub-grid-scale model (\Mycitep{daniele19a}; \Mycitep{carrasco20}; \Mycitep{daniele20}; \Mycitep{ricard20}; \Mycitep{palenzuela22}; \Mycitep{ricard22}; \Mycitep{leakage22}). Recently, we have also implemented the neutrino leakage scheme, with a novel formal calculation of the optical depth \Mycite{leakage22}. Now, we establish a new direction by incorporating the truncated moment formalism, including all the non-linear terms appearing in the matter-radiation interaction (see, e.g., \Mycitep{anninos20}; \Mycitep{rad2022}) and the  numerical methods presented in this work. 

The remainder of the paper is organized as follows. We summarize the truncanted moment formalism for radiation in Sec. \ref{S_truncated_moment_formalism}. We present the global time integrator, i.e., the IMEX schemes, the spatial discretization scheme, and the implicit treatment of the stiff source terms in Sec. \ref{S_general_framework}. We present the results of the numerical tests performed in Sec. \ref{S_numerical_tests}. Finally, conclusions and future work are discussed in Sec. \ref{S_conclusions}. For the reader only interested in the final IMEX schemes, they are summarized in Appendix \ref{A_summary_IMEX}. The comparison between the IMEX schemes designed in this work and two widely used semi-implicit schemes is presented in Appendix \ref{appendix_comparison_sec}. Unless otherwise noted, we employ a unit system in which $G=c=M_{\odot}=1$. We follow the usual convention according to which $\{a,b,c,d\}$ ($\{i,j,k\}$) indices refer to spacetime (spatial) coordinates and adopt a metric signature $(-,+,+,+)$.
%
\section{Truncated Moment Formalism} \label{S_truncated_moment_formalism}
The neutrino radiation transport evolves the distribution function of neutrinos in the 7-dimensional phase space $(x^a, p^a)$, where $x^a = (t, x^i )$ are the spacetime coordinates and $p^a$ are the components of the 4-momentum of the neutrinos, satisfying $p^a p_a= 0$ when the neutrino mass is neglected. The distribution function follows the Boltzmann equation, which depends not only of the space-coordinates, but also on the neutrino energies, species and propagation angles through the collision terms (i.e., absorption, emission and scattering of neutrinos). Solving the Boltzmann equation requires, therefore, the time evolution of a 6-dimensional function, a very demanding computational challenge even with modern facilities.

To overcome this issue we will adopt the moment formalism developed by (\Mycitep{thorne81}; \Mycitep{shi11}) in which only the lowest moments of the distribution function in momentum space are evolved. These radiation moment schemes are commonly referred to as M$n$ in the case where we evolve $n + 1$ moments. Recently, many works have considered only the first two moments of the distribution function (M1 scheme), where one evolves, for each species, projections of the neutrino radiation stress-energy tensor, $T^{ab}_{\text{rad}}$. Although the moment formalism can be used in theory with a discretization in neutrino energies, the additional dimension increases significantly the cost of the simulations. Morever, it involves additional technical difficulties in the treatment of the gravitational and velocity redshifts, particularly for applications such as compact binary mergers. Therefore, in order to reduce the computationally cost, we follow previous works and consider this formalism in the \emph{grey} approximation, i.e., evolving energy-integrated moments. 	

We will only consider three independent neutrino species: the electron neutrinos $\nu_e$, the electron antineutrinos $\bar{\nu}_e$, and the heavy-lepton neutrinos $\nu_x$. The latter is the combination of 4 species, $\nu_x = (\nu_{\mu} , \bar{\nu}_{\mu}, \nu_{\tau} , \bar{\nu}_{\tau})$. This merging is justified because the temperatures and neutrino energies reached in our merger calculations are low enough to suppress the formation of the corresponding heavy leptons, whose presence would require including the charged current neutrino interactions that differentiate between these individual species (see, e.g., \Mycitep{fou15}; \Mycitep{foucart_review}). The following is a brief summary of the formalism and the final evolution equations that describe each neutrino species.
%
\subsection{Neutrino Energy and Momentum Density Evolution: M1 approach}
In this section, we will describe the mathematical formalism following closely the approach presented in \Mycite{rad2022}. Let us consider a fluid with four-velocity $u^a$ moving on a spacetime described by a four-dimensional metric tensor $g_{ab}$. One possible decomposition of the neutrino radiation stress-energy tensor is given by
\begin{equation}
	T^{ab}_{\text{rad}} = J u^a u^b + H^a u^b + H^b u^a + Q^{ab}\,,
\end{equation}
where the radiation energy density $J$, the radiation flux $H^a$ and the radiation pressure tensor $Q^{ab}$ are evaluated by an observer co-moving with the fluid (i.e., the fluid-rest frame). Notice that, by construction, both the flux and the pressure tensors are orthogonal to the fluid velocity, i.e., $H^a u_a = Q^{ab} u_b = 0$.

It is also possible to decompose the same tensor $T^{ab}_{\text{rad}}$ with respect to a set of observers moving along $n^a$, the future-oriented unit normal to the $t = \text{const}$ hypersurfaces (i.e., the Eulerian frame), namely
\begin{equation}
	T^{ab}_{\text{rad}} = E n^a n^b + F^a n^b + F^b n^a + P^{ab}\,,
\end{equation}
where the new quantities $E$, $F^a$ and $P^{ab}$ appearing in this decomposition are respectively the radiation energy density, the radiation flux, and the radiation pressure tensor, evaluated in the Eulerian frame. Again, note that $F^{a}$ and $P^{a b}$ are purely spatial by construction, $F^a n_a = P^{ab} n_b = 0$ (i.e., $F^0 = P^{0a} = 0$ ). 

We can express the Eulerian quantities, $\{ E, F_a, P_{ab}\}$,  in terms of those of the fluid-rest frame, $\{ J, H^a, Q^{a b}\}$, by decomposing the fluid four-velocity as $u^{a} = W(n^{a} + v^{a})$, where $W$ is the Lorentz factor and $v^{a}$ the spatial velocity of the fluid,
\begin{align}
	E &= W^2 J + 2W v_{a} H^{a} + v_{a} v_b Q^{a b}\,, \\
	F_{a} &= W^2 v_{a} J + W(g_{ab} - n_{a} v_b) H^b \\
	& \quad + (g_{ab} - n_{a} v_b)v_c Q^{bc} + W\,v_{a} v_b H^b\,, \nonumber \\
	P_{ab} &= W^2 v_{a} v_b J + W(g_{ac} - n_{a}
	v_c)v_b H^c \\
	&\quad+ W (g_{cb} - n_b v_c)v_{a} H^c \nonumber \\
	&\quad + (g_{ac} - n_{a} v_c)(g_{bd} - n_b
	v_d)Q^{cd}\,.\nonumber
\end{align}
Conversely, the quantities of the fluid-rest frame can be obtained from the Eulerian ones by means of
\begin{align}
	J &= W^2 (E - 2 F^{a} v_{a} + P^{ab} v_{a}v_b)\,,  \\
	H^{a} &= W( E - F^b v_b)h^{a}_{~c} n^c + W h^{a}_{~b} ( F^b - v_c P^{bc})\,, \\
	Q^{ab} &= T^{cd}_{\rm{rad}}\,h^{a}_{~c}
	h^b_{~d}\,,
\end{align}
where $h_{ab} \equiv g_{ab} + u_{a} u_b$ is the projection tensor orthogonal to the four-velocity of the fluid, $h_{ab} u^{a}=0$. 
Notice that when the fluid is at rest, $v_i = 0$, the translation between frames is trivial, i.e., $E=J$, $F^{a} = H^{a}$ and $P^{a b} = Q^{a b}$.

Conservation of energy and angular momentum implies
\begin{equation}
	\nabla_b T_{\text{rad}}^{ab} = {\cal S}^a\,, \label{eq_conservation}
\end{equation}
where $\nabla$ is the covariant derivative operator compatible with the spacetime metric $g_{a b}$ and ${\cal S}^a$ is the term representing the interaction between the neutrino radiation and the fluid, which can be written as
\begin{equation}
{\cal S}^a = (\eta - \kappa_a J) u^a - (\kappa_a + \kappa_s) H^a\,,
\end{equation}
where $(\eta,\kappa_a,\kappa_s)$ are respectively the (energy-averaged) neutrino emissivity, absorption opacity and scattering opacity.

These covariant equations can be written explicitly as evolution equations by performing the $3+1$ decomposition. First, the spacetime metric is decomposed as
\begin{equation}
ds^2 = -\alpha^2\,dt^2 
+ \gamma_{ij}\left(dx^i + \beta^i\,dt\right)\left(dx^j + \beta^j\,dt\right)\,,
\end{equation}
where $\alpha$ is the lapse function, $\beta^{i}$ the shift vector, $\gamma_{ij}$ the induced 3-metric on each spatial slice, and $\sqrt{\gamma}$ is the square root of its determinant. Within this decomposition, the normal vector to the hypersurfaces is just $n_a = (-\alpha, 0)$. It is also standard to define the extrinsic curvature $K_{ij}$ as the Lie derivative along this normal vector.

The conservation (eq. \ref{eq_conservation}) in the $3+1$ decomposition \Mycite{shi11} can be written as
\begin{eqnarray}
	&\partial_t (\sqrt{\gamma} E)&  \label{eq_evolE}
	 + \partial_i \left[ \sqrt{\gamma} (\alpha F^i - \beta^i E)  \right] = \\
    & & \alpha \sqrt{\gamma} \left[ P^{ij} K_{ij}
	- F^i (\partial_i \alpha)/\alpha - {\cal S}^a n_a  \right]\,, \nonumber \\
	&\partial_t (\sqrt{\gamma} F_i)& \label{eq_evolF}
	 +\partial_j \left[ \sqrt{\gamma} (\alpha {P^j}_{i} - \beta^j F_i)  \right] = \\
	& & \sqrt{\gamma} \left[ - E \partial_i \alpha + F_j \partial_i \beta^j + \frac{\alpha}{2} P^{kj} \partial_i \gamma_{kj} + \alpha {\cal S}^a \gamma_{ia}  \right]\,.   \nonumber
\end{eqnarray}	
which strongly resemble the hydrodynamical equations except by the fluid-neutrino interaction term.
%
\subsection{\textbf{Algebraic Closure for the M1}}
The evolution equations (eqs. $\ref{eq_evolE}$, $\ref{eq_evolF}$) are exact but do not have closed form, since in general $P^{ij}$ depends on the global geometry of the fields $(E,F^k)$. Therefore, only approximate closures can be introduced for these equations when using the local values of $(E,F^k)$. There is no single closure prescription $P^{ij}=P^{ij}(E,F^k)$ that describes accurately all the possible regimes. Instead, the M1 scheme introduces an analytic closure that focuses on those limits where well-tested ansatzs are properly defined (\Mycitep{rad2022}). For a more comprehensive description of the limitations of the M1 scheme, we refer interested readers to (\Mycitep{gavassino20}; \Mycitep{sado13}).

is point and its theoretical implications (it defines the limitations of the M1 scheme) are clearly addressed
Then, the M1 scheme is constrained by two limits: the \emph{optically thick limit} (where matter and radiation are in thermodynamic equilibrium) and the \emph{optically thin limit} (propagation of radiation in a transparent medium from a point source). Fortunately, we can obtain explicit expressions for $P_{ij}$ in both limits, $\{P_{ij}^\text{thin}, P_{ij}^\text{thick} \}$, and construct the following expression
\begin{equation}\label{eq_Ptot}
P_{ij} = \frac{3 \chi(\xi) - 1}{2} P^{{\text{thin}}}_{ij} + \frac{3 [1- \chi(\xi)]}{2} P^{{\text{thick}}}_{ij}\,,
\end{equation}
where $\chi(\xi) \in \left[\frac{1}{3}, 1\right]$ is the \textit{closure function} and $\xi$ is the \textit{variable Eddington-factor}. This equation is derived based on the assumption that the radiation is symmetric about the direction parallel to the flow. Although this assumption is valid for spherically symmetrical matter and radiation distributions, it is not always true in more general cases. For instance, in the optically thin limit, the collision of radiation beams from different sources will lead to the non-physical behavior of shock formation. 

As discussed in \Mycite{shi11}, there are many possible choices for $\xi$, but the only one that is accurate in the optically thick limit is
\begin{equation} \label{eq_xi}
		\xi \equiv \sqrt \frac{H_a H^a}{J^2}\,,
\end{equation}
where $H_a H^a = H_k H^k - H_n^2$. Unfortunately, this choice is computationally expensive, since the calculation of $\xi$ requires a root-finding method to transform the fields from the Eulerian to the fluid-rest frame.

Regarding the \textit{closure function}, $\chi$, we choose the commonly employed \textit{Minerbo closure} \Mycite{minerbo79}, which is exact in both limits,
\begin{equation}\label{eq_minerbo}
\chi (\xi)= \frac{1}{3} + \xi^2 \left(
\frac{6 - 2\xi + 6 \xi^2}{15} \right)\,.
\end{equation}
Notice however that there are many other choices that might also capture accurately these two limits (for more details see \Mycitep{mur17}).
%
\subsubsection{\textbf{Optically Thin Limit ${(\xi = 1 )}$}}
In the optically thin limit we assume that radiation is streaming at the speed of light in the direction of the radiation flux, leading to the explicit relation
\begin{equation}\label{eq_Pthin}
	P^{{\text{thin}}}_{ij} = \frac{F_i F_j}{F^k F_k} E \,,
\end{equation}
In these regions, $H^a  \approx J$ (i.e., $\xi \approx 1$ and $\chi \approx 1$), so $P_{ij} \approx P^{{\text{thin}}}_{ij}$.
Unfortunately, this limit is not unique, as it is determined by the non-local geometry of the radiation field. In general, it will not correctly describe free-streaming neutrinos, since in vacuum they usually do not propagate in the same direction. For instance, the trajectory of colliding beams will propagate unphysically in the direction of their average momentum, as it is very well represented by the well-known ``colliding beams'' test (see, e.g., \Mycitep{sado13}; \Mycitep{fou15}; \Mycitep{weih20}). This behavior, in addition to being responsible for radiation shocks, can lead to incorrect results in some scenarios (\Mycitep{rad2022}).
%
\subsubsection{\textbf{Optically Thick Limit ${(\xi = 0 )}$}}
In the optically thick limit (or diffusion limit), the radiation pressure tensor measured by the fluid-rest frame is described by
\begin{equation}
	Q^{{\text{thick}}}_{ab} = \frac{J_{{\text{thick}}}}{3} (g_{ab} + u_a u_b)\,.
\end{equation}
To obtain the tensor, $P^{{\text{thick}}}_{ij}$, in the Eulerian frame, we are forced to compute the intermediate quantities, $\{J_{\text{thick}}, H^{i}_{\text{thick}}, H^{n}_{\text{thick}}\}$, that relate the fluid-rest frame with the Eulerian frame, namely
\begin{eqnarray}
	&&J_{{\text{thick}}} = \frac{3}{2 W^2 + 1} \left[ (2 W^2 - 1) E - 2 W^2 F^k v_k \right]\,, \label{eq_Jthick} \\
	&&H^n_{{\text{thick}}} = -H^a n_a = W ( E - J_{{\text{thick}}} - F^k v_k) \,,\\
	&&H^i_{{\text{thick}}} = \gamma^i_b H^b_{{\text{thick}}} = \frac{F^i}{W} - \frac{4}{3} J_{{\text{thick}}} W v^i - H^n_{{\text{thick}}} v^i  \label{eq_Hithick} \\
	&&~~~~=  \frac{F^i}{W} + \frac{W v^i}{2 W^2 + 1} 
	\left[(4 W^2 + 1) F^k v_k - 4 W^2 E \right]\,, \nonumber \\
	&&P_{ij}^{{\text{thick}}} = \frac{4}{3} J_{{\text{thick}}} W^2 v_i v_j \nonumber \\
	&&~~~~	+ W (H^{{\text{thick}}}_i v_j + H^{{\text{thick}}}_j v_i) + \frac{1}{3} J_{{\text{thick}}} \gamma_{ij} \,.
	\label{eq_Pthick}
\end{eqnarray}
In the optically thick regions the radiation satisfies $H^a  \approx 0$ (i.e., $\xi \approx 0$ and $\chi \approx 1/3$), so $P_{ij} \approx P^{{\text{thick}}}_{ij}$. 
%
\subsubsection{\textbf{Minerbo Closure  $\left({\xi} \in [0,1]\right)$}}
As we have already discussed, to close the evolution equations (eqs. $\ref{eq_evolE}$, $\ref{eq_evolF}$) for $(E, F^i)$ we need to find $\xi$ which depends on $(J, H^{a})$, which in turn depends on the unknown $P_{ij}$. Therefore, it is necessary to calculate $\xi$ using a root-finding method. First, we need to elucidate the relationship between the frames. 
By considering the definitions presented in Sec. \ref{S_truncated_moment_formalism}, we can arrange these relations as follows
\begin{eqnarray}\label{eq_shibataJ}
	-J W + H^a n_a = - E W + W F^k v_k \,, \\
	\label{eq_shibataH}
	J W v_i + H_i = W F_i - W P^k_i v_k \,,
\end{eqnarray}
such that it is possible to reconstruct $H^a = - (H^b n_b) n^a + \gamma_k^a H^k$, with $-H^a n_a = H^n$ and $\gamma_{ai} H^a = \gamma_{ai} \gamma_k^a H^k = H_i$. Thus, the idea is that at each point one can compute $P_{ij}$ using the Minerbo closure (eq. \ref{eq_Ptot}) and the evolved fields $\{E,F^i\}$. However, since $\chi (\xi (J, H^a))$ as described by (eq. \ref{eq_minerbo}), and those fields depend also on $P_{ij}$ through (eqs. \ref{eq_shibataJ}, \ref{eq_shibataH}), we obtain a transcendental equation which needs to be solved numerically. The function to be solved is
\begin{equation}\label{eq_minerbo_transcendental}
	R = \frac{\xi^2 J^2 - H_a H^a}{E^2} = 0 \,,
\end{equation}
for which we need to compute the derivatives, either analytically or numerically, of this function. We perform the root-finding by adopting the \emph{Van Wijngaarden-Dekker-Brent} method from numerical recipes in \texttt{C++} \Mycite{recipesC}, which uses a combination of root bracketing, bisection and inverse quadratic interpolation to converge from the neighborhood of a zero crossing. This method is guaranteed to converge as long as the function can be evaluated within the initial interval known to contain a root and does not require analytical knowledge of derivatives. The numerical recipe would be:
\begin{enumerate}
	\item Compute $\{P_{ij}^{{\text{thin}}},P_{ij}^{{\text{thick}}}\}$ following (eqs. \ref{eq_Pthin} - \ref{eq_Pthick}).
	\item Using as an initial guess for $\chi$ the value from the previous step, compute the total $P_{ij}$ using (eq. \ref{eq_Ptot}).
	\item Calculate the components of $H^a$ following these relations
	\begin{eqnarray}\label{eq_Ha_procedure}
		J &=& W^2 (E - 2 F^k v_k + P^{ij} v_i v_j) \,,\\
		H_n &=& - H^a n_a = W (E - J - F^k v_k) \,, \\
		H_i &=& W (F_i - J v_i - {P^k}_i v_k) \,.	
	\end{eqnarray}
	\item Compute $H_a H^a = H_k H^k - H_n^2$ and $\{\xi,\chi\}$ using (eqs. \ref{eq_xi}, \ref{eq_minerbo}).
	\item Check if $R=0$ as given by (eq. \ref{eq_minerbo_transcendental}). Otherwise, use these values to compute a new guess for $\chi$ using the \emph{Van Wijngaarden-Dekker-Brent} method and iterate through the steps 2--4 until convergence.
\end{enumerate}
%
\subsection{Neutrino Number Density Evolution: N0 approach}
To obtain valuable information of the neutrino energies without evolving an energy-dependent scheme, it is useful to evolve the number density of neutrinos, $n$. Weak interactions preserve the total lepton number of the system, leading to a non-conserved evolution equation for the electron fraction of the matter $Y_e$ (i.e., the ratio of electrons to baryons, defined in eq.~\ref{Ye}). Calculating accurately the evolution of $Y_e$ in  BNS merger simulations is one of the main reasons to include neutrino transport and the detailed neutrino-matter interactions. Following Foucart's approach (\Mycitep{fou16b}; \Mycitep{foucart_review}; \Mycitep{rad2022}), for each neutrino species we introduce a neutrino number current $N^a$ following a conservation equation 
\begin{equation}
\nabla_a N^a = {\cal S}_N =  \eta^0 - \kappa_a^0 n \,,
\end{equation}
where $n = -N^a u_a$ is the neutrino density in the fluid-rest frame and $(\kappa_a^0, \eta^0)$ are the neutrino number absorption and emission coefficients, also to be computed from the fluid state and the information in the equation of state (EoS) tables.

Assuming that the neutrino number density and the radiation flux are aligned (i.e., which is a reasonable assumption but not true in general), one can define the following closure relation
\begin{equation}
N^a = n f^a = n \left(u^a + \frac{H^a}{J}\right) \,,
\end{equation}
which we call N0, as only the first moment of the neutrino number density is evolved.

The evolution of the neutrino number density within the $3+1$ decomposition can then be written as
\begin{equation}
\partial_t (\sqrt{\gamma} n \Gamma) + 
\partial_i (\alpha \sqrt{\gamma} n f^i)
= \alpha \sqrt{\gamma} (\eta^0 - \kappa_a^0 n)  
\end{equation}
where
\begin{eqnarray}
\Gamma &\equiv& \alpha f^0 = W - \frac{1}{J} H^a n_a = 
W \left( \frac{E - F_a v^a}{J} \right)
\,, \\
f^i &\equiv& W \left(v^i - \frac{\beta^i}{\alpha}\right) + \frac{H^i}{J} \,.
\end{eqnarray}

With the $\text{N}0+\text{M}1$ approach one can compute the average energy of the neutrinos $\epsilon_{\nu}$, since in the fluid-rest frame approximately $J = n \langle \epsilon_{\nu} \rangle$. By defining the evolved field $N = n \Gamma$, one can write then
\begin{equation}
\langle \epsilon_{\nu}\rangle = \frac{J}{n}
= \frac{\Gamma J}{N} = \frac{W (E - F^i v_i)}{N} \,,
\end{equation}
which gives a way to calculate an average energy that varies with space and time.
%
\section{General Framework}\label{S_general_framework}
\subsection{Evolution Equations} \label{SS_M1_evol_eqs}
The covariant set of equations that models a self-gravitating magnetized fluid with neutrinos includes Einstein equations, in which the geometry of the spacetime is described by the Einstein tensor, $G_{ab}$, coupled to the stress-energy tensor which describes matter, $T_{ab}$. Here we consider the matter fields to be a combination of the perfect fluid, $T_{a b}^{\text{fl}}$, and the neutrino radiation, $T_{a b}^{\text{rad}}$, with three independent neutrino species $\nu= (\nu_e, \bar{\nu}_e, \nu_x)$. Therefore, Einstein equations can be written as:
\begin{eqnarray}
	G_{ab} &=& 8\pi T_{a b} = 8\pi \left(T_{a b}^{\text{fl}} + T_{ab}^{\text{rad}} \right) \,, \\
	T_{ab}^{\text{rad}} &=& \sum_{\nu} T_{ab}^{\nu}
	= T_{ab}^{\nu_e} + T_{ab}^{\bar{\nu}_e} + T_{ab}^{\nu_x} \,.
\end{eqnarray}

The dynamics of the matter are described by a series of (quasi-)conservation laws for:
\begin{itemize}
	\item The energy and momentum densities of each neutrino species, namely
	\begin{eqnarray}\label{neutrino_eqs}
		\nabla_{a}T^{a b}_{\nu} &=&  \mathcal{S}_{\nu}^{b} \,,
	\end{eqnarray}
	which can be written explicitly as evolution equations by using the $3+1$ decomposition described in the previous section.	
	\item The equations for a magnetized fluid follow from the energy-momentum conservation of the total stress-energy tensor $T_{ab}$, and the conservation of the Faraday tensor, $F^{ab}$,  namely
	\begin{eqnarray}\label{1fluid_eqs}
		\nabla_{a}T^{a b}_{\text{fl}} &=& - \sum_{\nu} \mathcal{S}^{b}_{\nu} 
		= - \left(  \mathcal{S}^{b}_{\nu_e} + \mathcal{S}^{b}_{\bar{\nu}_e} + \mathcal{S}^{b}_{\nu_x} \right) \,, \\ \label{2fluid_eqs}
		\nabla_{a}\text{*} F^{ab} &=& 0 \,,
	\end{eqnarray}
	which, with the $3+1$ decomposition, can be written as the standard MHD equations with an extra source term arising from the interaction of the fluid with the neutrinos.
	\item The conservation of baryonic number, neutrino number and lepton number densities,
	\begin{eqnarray}
		\nabla_{a}(\rho u^{a}) &=& 0 \,, \\
		\nabla_{a} N_{\nu}^{a} &=& {\cal S}^N_{\nu} \,,		\\	
		\nabla_{a}(Y_{e}\rho u^{a}) &=& m_b \alpha \left( {\cal S}^N_{\bar{\nu}_e} - {\cal S}^N_{\nu_e} \right) \,,
	\end{eqnarray}
	where, $\rho$ is the rest-mass density, $Y_e$ is the electron fraction and $m_b$ the reference baryon mass. They are related by the following equations:
	\begin{equation}\label{Ye}
		\rho = m_b (n_p + n_n) \,,~~~ Y_e = \frac{n_{e_-} - n_{e_+} }{n_p + n_n} \,,
	\end{equation}	
	where $n_i$ is the number density of species $i$ in the fluid frame. 
\end{itemize}
%
\subsection{Hyperbolic-Diffusion Equations}
The evolution fields, corresponding to the previous equations, include: the spacetime, the magnetized fluid, and the neutrino radiation fields. Both Einstein and MHD equations are hyperbolic, meaning that the information propagates with finite speed. However, the M1 equations for the neutrinos contain potentially stiff terms, resulting in a more general class of advection-diffusion systems. These system of equations with stiff terms can be solved by employing the Method of Lines (MoL), for which is useful to decompose the equations as
\begin{eqnarray}\label{eq_stiff}
	\partial_t \textbf{U} &=&  \mathcal{F} (\textbf{U})  + \frac{1}{\epsilon} \mathcal{R} (\textbf{U}) \,,
\end{eqnarray}
where  $\mathcal{R} (\textbf{U})$ accounts for the stiff part (i.e., the neutrino-fluid interaction terms in the neutrino radiation transport equations) while $\mathcal{F}(\textbf{U})$ accounts for all the other non-stiff terms which can be treated explicitly (i.e., fluid and general relativistic terms). Here, $\epsilon$ is the relaxation time. It is assumed that in the limit $\epsilon \rightarrow \infty$ the system is hyperbolic with a spectral radius $c_h$ (i.e., the absolute value of the maximum eigenvalues). At the other limit $\epsilon \rightarrow 0$, the system is clearly stiff since the time scale $\epsilon$ of the relaxation (or stiff term) $\mathcal{R} (\tmmathbf{U})$ is very different from the speeds $c_h$ of the hyperbolic (or non-stiff) part $\mathcal{F} (\tmmathbf{U})$. In the stiff limit ($\epsilon \rightarrow 0$) the stability of an explicit time evolution scheme \Mycite{But1987} is only achieved with a time step size $\Delta t \leq \epsilon$. This restriction is much stronger than the one given by the Courant-Friedrichs-Lewy (CFL) condition $\Delta t \leq \Delta x/c_h$. 

The dynamical behavior of the neutrinos is determined by evolution equations (eq. \ref{neutrino_eqs}) that may exhibit stiffness due to the presence of source terms that  depend explicitly on the evolution fields. As noted in \Mycite{foucart_review}, during the evolution the coupling between neutrinos and matter in certain regions can give rise to additional stiff source terms that influence the fluid evolution equations (eq. \ref{1fluid_eqs}). These stiff terms are extremely challenging to solve implicitly due to the complex dependence of the fluid-neutrino interaction term. For this reason, following previous works (see, e.g., \Mycitep{kuroda16}; \Mycitep{weih20}; \Mycitep{rad2022}; \Mycitep{zappa22}; \Mycitep{foucart_review}), we solve only implicitly the source terms of the neutrino moment equations, while handling explicitly the source terms of the fluid equations. Consequently, due to the distinct characteristic structure of these equations, we split the vector fields, $\bf{U}$, into two parts, i.e., $\bf{U} = (\bf{V},\bf{W})$, where $\bf{V}$ will gather the neutrino fields, which contain stiff terms, while $\textbf{W}$ includes the fluid and spacetime fields which only have non-stiff right-hand-sides. The equations can then be written
\begin{eqnarray}\label{eq_split}
 	\partial_t {\bf W} &=& F_W({\bf V},{\bf W}) \,, \\
	\partial_t {\bf V} &=& F_V({\bf V},{\bf W}) + \frac{1}{\epsilon} R_V({\bf V},{\bf W}) \,.
\end{eqnarray}

The development of efficient numerical schemes for such systems is challenging, since in many applications the relaxation time can vary by many orders of magnitude, from order one to very small values compared to the time scale determined by the characteristic speeds of the system. There are several ways to deal with hyperbolic systems with relaxation or diffusion terms.  These different time scales could be solved using an implicit scheme, but its computational cost in 3D makes it unfeasible to use it alone for all the equations. To overcome this problem, several approaches have been developed in recent years in the context of the truncated moment formalism for neutrino transport.

One possibility is to bound the stiff terms in such a way that no implicit steps are needed \Mycite{sun22}. This approach, while greatly simplifies the numerical simulation, is only valid for short timescale and can lead to non-realistic results. In contrast, other authors employ a semi-implicit scheme that combines an implicit and an explicit step (see, e.g., \Mycitep{rampp02}; \Mycitep{vaytet11}; \Mycitep{sado13}; \Mycitep{rad13}; \Mycitep{nagakura14}; \Mycitep{just15}; \Mycitep{kuroda16}; \Mycitep{skinner19}; \Mycitep{chan20}; \Mycitep{rad2022}). Although they are well-known to the community and fairly easy to implement, they also have two important issues. First, the low effective CFL and the low accuracy (usually, it is only first order) of the scheme might degrade the solution quickly. Second, the non-stiff Einstein and MHD equations do not need any implicit time integration, so they are usually evolved with higher-order Runge-Kuttas. This splitting of the evolution integration schemes involves some time interpolation between the two sets of fields $(\bf{V},\bf{W})$, which could further degrade the accuracy and the efficiency of the integration scheme. 

In this paper, we present a different approach to solve these issues and limitations, by using IMEX RK schemes (see, e.g., \Mycitep{ascher97}; \Mycitep{pareschi00}; \Mycitep{kennedy03}; \Mycitep{pareschi05}; \Mycitep{boscarino07}). These methods have been employed in Numerical Relativity (NR) in the context of resistive magneto-hydrodynamics (see, e.g., \Mycitep{palenzuela09}; \Mycitep{alic12}; \Mycitep{dio13}; \Mycitep{ripperda19}). More recently, they have been also used in the neutrino radiation problem (see, e.g., \Mycitep{mck14}; \Mycitep{chu19}; \Mycitep{weih20}; \Mycitep{laiu21}), to substitute the semi-implicit method. With them, one can use the same generic integration scheme, where only the stiff terms are treated implicitly, while keeping high-order accuracy (at least second order). As we will see later, we will enforce the condition that the explicit RK methods are exactly the same ones commonly used in NR applications. This has two important advantages: first, we can still use high-order schemes with a large effective CFL factor, and second, it minimizes the errors during sub-cycling in time when using Fixed Mesh Refinement (FMR) and AMR. We will discuss these issues in detail in the next subsection.
%
\subsection{Temporal Discretization: IMEX Schemes}\label{SS_time_integration_section}
An IMEX Runge-Kutta scheme consists of applying an implicit discretization to the stiff terms and an explicit one to the non-stiff ones. When applied to the system (eq. \ref{eq_stiff}) it takes the form
\begin{eqnarray}\label{eq_IMEX}
	{\bf U}^{(i)} = {\bf U}^n &+& \Delta t \sum_{j=1}^{i-1} {\tilde{a}}_{ij} \mathcal{F}({\bf U}^{(j)}) 
	\nonumber \\
	&+& \Delta t  \sum_{j=1}^{i} a_{ij} \frac{1}{\epsilon} \mathcal{R}({\bf U}^{(j)}) \,, \\
	{\bf U}^{n+1} = {\bf U}^n &+& \Delta t \sum_{i=1}^{q} {\tilde{\omega}}_{i} \mathcal{F}({\bf U}^{(i)})
	+ \Delta t  \sum_{i=1}^{q} \omega_{i} \frac{1}{\epsilon} \mathcal{R}({\bf U}^{(i)}) \,,
	\nonumber
\end{eqnarray}
where ${\bf U}^{(i)}$ are the auxiliary intermediate values of the Runge-Kutta. The matrices $\tilde{A}= (\tilde{a}_{ij})$, $\tilde{a}_{ij} = 0$ for $j \geq i$ and $A= (a_{ij})$ are $q \times q$ matrices such that the resulting scheme is explicit in $\mathcal{F}$ and implicit in $\mathcal{R}$ \Mycite{pareschi05}. An IMEX Runge-Kutta is characterized by these two matrices and the coefficient vectors $\tilde{\omega}_i$ and $\omega_i$. Since the simplicity and efficency of solving the implicit part at each step is of great importance, it is natural to consider diagonally-implicit Runge-Kutta (DIRK) schemes ($a_{ij}=0$ for $j > i$) for the stiff terms.
IMEX RK schemes can be represented by a double \emph{tableau} in the usual Butcher notation (\Mycitep{But1987})
\begin{equation}
	\begin{minipage}{1in}
		\begin{tabular} {c c c}
			${\tilde c}$  & \vline & ${\tilde A}$  \\
			\hline 
			& \vline & ${\tilde \omega}^T$ \\
		\end{tabular}
	\end{minipage}
	\begin{minipage}{1in}
		\begin{tabular} {c c c}
			${c}$  &  \vline & ${A}$  \\
			\hline 
			&  \vline & ${\omega}^T$  \\
		\end{tabular}
	\end{minipage}
	\label{butcher_tableau}
\end{equation}
where the coefficients $\tilde{c}$ and $c$ used for the treatment of non-autonomous systems are given by the following relations
\begin{eqnarray}\label{eq_definition_cs}
	{\tilde c}_{i} = \sum_{j=1}^{i-1}~ {\tilde{a}}_{ij} \,, ~~~~~~
	{c}_{i} = \sum_{j=1}^{i}~ {a}_{ij}.
\end{eqnarray}

The notations we adopt are $(\tilde A, \tilde w, \tilde c)$ and $(A,w,c)$ to characterize the explicit and implicit Butcher \emph{tableau}, respectively. The IMEX RK schemes will be denoted as RK($s, \sigma, p$), where the triplet ($s, \sigma, p$) characterizes the number $s$ of stages of the implicit scheme, the number $\sigma$ of stages of the explicit scheme and the order $p$ of the IMEX scheme.

These RK schemes need to satisfy the following conditions to achieve high order accuracy, \\
\textbf{First order}
\begin{equation}
	\sum_{i}^q \tilde{w}_i = 1 \,, \quad \sum_{i}^q w_i = 1 \,,
	\label{eq_RKcond1}
\end{equation}
\textbf{Second order}
\begin{equation}
	\sum_{i} \tilde{w}_i \tilde{c}_i = 1/2 \,, \quad \sum_i w_i c_i = 1/2 \,,
	\label{eq_RKcond2}
\end{equation}
\textbf{Third order}
\begin{eqnarray}
	\sum_{ij} \tilde{w}_i \tilde{a}_{ij} \tilde{c}_j = 1/6\,,~~~~
	\sum_{i}  \tilde{w}_i \tilde{c}_{i}  \tilde{c}_i = 1/3\,,\label{eq_RKcond3_e} \\
	\sum_{ij}   w_i   a_{ij}   c_j = 1/6\,, ~~~~
	\sum_{i}    w_i   c_{i}    c_i = 1/3\,.
	\label{eq_RKcond3}
\end{eqnarray}
From these equations we can conclude that the characteristics of these schemes will be quite similar to those of an explicit RK. Apart from other subtleties, the major difference will come when constructing a high order IMEX scheme. The derivation of the order conditions of RK schemes is based on the Taylor expansion of the exact and numerical solution. Unfortunately, the coupling of the implicit and explicit Butcher \emph{tableau} gives rise to several additional conditions that must be fulfilled in order to satisfy a given order of accuracy. Pareschi and Russo \Mycite{pareschi05} summarized the number of coupling conditions of IMEX RK schemes (see Table \ref{conditionsIMEX}), from 
\begin{table}
	\begin{center}
		\begin{tabular}{c|c|c|c|c}
			{\bf IMEX-RK} & \multicolumn{4}{c}{\bf Number of coupling conditions}\\
			{\bf order}  & General case  & $\tilde{w}_i = w_i$ & $\tilde{c} = c$ &
			$\tilde{c} = c$ and $\tilde{w}_i =
			w_i$\\
			\hline
			1 & 0 & 0 & 0 & 0\\
			2 & 2 & 0 & 0 & 0\\
			3 & 12 & 3 & 2 & 0\\
			4 & 56 & 21 & 12 & 2\\
			5 & 252 & 110 & 54 & 15\\
			6 & 1128 & 528 & 218 & 78
		\end{tabular}
	\end{center}
	\caption{Number of Coupling Conditions in IMEX Runge-Kutta Schemes \Mycite{pareschi05}. \label{conditionsIMEX}}
\end{table}
where it is easy to conclude that in general it is difficult to find a high order IMEX scheme. 
%
\subsubsection{\textbf{Our Requirements}}\label{SSS_IMEX_requirements}
After reviewing the IMEX schemes, we need to address what are the desired features in our simulations. It is important to remember that the results presented in this paper describe a general framework for accurately modeling the influence of neutrinos after the merger of a BNS system. Solving the stiff term, $\mathcal{R}(\mathbf{U})$, is computationally expensive in the neutrino radiation transport problem, so we will try to avoid it as much as possible.  Besides that, we will use AMR with sub-cycling in time (see e.g., \Mycitep{berger84}; \Mycitep{berger89}), such that at the boundaries between refinement levels we need to interpolate the solution from the coarse mesh in order to fill the boundary data for the fine mesh solution.
There is an efficient and accurate treatment of the refinement boundaries that uses a minimal stencil and achieves high-order (\Mycitep{mccor11}; \Mycitep{mon15}), but imposes further conditions on the time discretization scheme. 

In summary, the complete list of requirements would be:  
\begin{enumerate}
	\item Maintain the explicit RK that is usually employed to integrate the non-stiff evolution equations (i.e., fluid and spacetime). The explicit RK is chosen such that achieves high order with the largest stable CFL factor. In addition, the explicit RK is usually  Strong-Stability-Preserving (SSP) (see, e.g., \Mycitep{shu_osher}; \Mycitep{shu88b}; \Mycitep{gottlieb98}; \Mycitep{gottlieb01}; \Mycitep{gottlieb03}) in order to deal with shocks appearing in the solutions. Notice that with this choice the non-stiff equations remain unchanged.
	
	\item Achieve the highest order of convergence with the minimum number of stages in the implicit RK. We also want to maximize the effective CFL, which can be defined as $\text{CFL}_{\text{eff}} = \text{CFL}/s$, of the IMEX, being $s$ the number of RK substeps (stages).
	
	\item The scheme must be able to deal effectively with problems presenting multiple scales. This results in a stability condition that is not constrained by the small scales of the system. Obviously, this is specially important to deal efficiently with the stiff limit $\epsilon \rightarrow 0$.
	
	\item Both the explicit and implicit RK must be evaluated at the same time (i.e.,  $\tilde c_i = c_i$) to avoid interpolation between the stiff and the non-stiff evolved fields at each RK substep. \label{c_equal_tilde_c}
	
	\item A high-order accuracy of the IMEX is desired, which implies satisfying not only the order conditions for each RK Butcher \emph{tableau}, but also the coupling order conditions. Choosing  $\mathbf{\tilde{\omega}}_i ={\omega}_i$ trivially satisfies the second order and some third order coupling conditions. \label{w_equal_w}
\end{enumerate}

Let us remind some basic concepts on time integration numerical schemes. The stability region for a generic equation $\dot{y} = \lambda y$ is the set of points such that $|\mathcal{R}(\lambda \Delta t )|<1$, where the stability function is given by \Mycite{hairer00}
\begin{equation} \label{eq_stability_function_RK}
\mathcal{R}(z)=1+zw^{T}(I-zA)^{-1}\mathbf{1} ~,
~~\mathbf{1}=(1,...,1)^{T} \,.
\end{equation}
Unconditional stability is guaranteed if the implicit RK is asymptotically preserving \Mycite{shi99}, meaning that: 
\begin{enumerate}
	\item The method is $A(\alpha)$-stable (i.e., there is a cone on the negative real semi-plane of angle $0< \alpha \leq \pi/2$ where the method is absolutely stable).
	\item The modes are strongly damped for large values of $z$, namely
	\begin{equation} \label{eq_stability_function_RK_infinity}
		\lim\limits_{z\rightarrow \infty}
		\mathcal{R}(z)=0 \,.
	\end{equation}
\end{enumerate}
The methods that satisfy these two conditions are called $L(\alpha)$-stable. The $A(\alpha)$-stable case with $\alpha=\pi/2$ (i.e., meaning that the stability region includes the entire left half-plane $z < 0$) is called $A$-stable, which becomes L-stable if the condition (eq.~\ref{eq_stability_function_RK_infinity}) is also satisfied \Mycite{piao18}

A consequence of the condition $\tilde c_i = c_i$ (cond. [\ref{c_equal_tilde_c}]) is that  $a_{11}=0$. These implicit schemes are known as Carpenter-Kennedy (CK) type \Mycite{kennedy03}. Then, such schemes also allow us to reduce the number of implicit steps. The CK implicit \emph{tableau} can be written in the following form
\begin{table}[h!]
	\begin{minipage}{1.4in}
		\begin{tabular} {c| c c}
			$0$ & $0$ & $0$ \\
			$\hat{c}$ & $a$ & $\hat{A}$ \\  
			\hline \\ [-1.0em] 
			$~$ & $w_1$& $\hat{w}^T$ \\
		\end{tabular}
	\end{minipage}
\end{table} \\
where it is assumed that the sub-matrix $\hat{A}$ is invertible (i.e., or
equivalently $a_{ii} \neq 0$ for $i = 2,..s$).

Here we consider at least $L(\alpha)$-stable, stiffly accurate implicit schemes (i.e., $w_i=a_{si}$, $i=1,\ldots,s$).
Notice that for the CK-type schemes, the stiffly accurate condition 
alone
\begin{equation}
	\hat{e}^T\hat{A}=\hat{w}^T \,, 
\end{equation}
where $\hat{e}=(0,0,1)^T$, does not guarantee the damping of modes at large $z$ (eq. \ref{eq_stability_function_RK_infinity}). In addition, one needs to satisfy the relation \Mycite{bos09}
\begin{equation} \label{eq_additional_cond_RK_infinity}
	\hat{e}^T\hat{A}^{-1} a = 0 \,, 
\end{equation}
as we will see in detail below.

\begin{figure}[t!]
	\centering
	\includegraphics[width=\columnwidth]{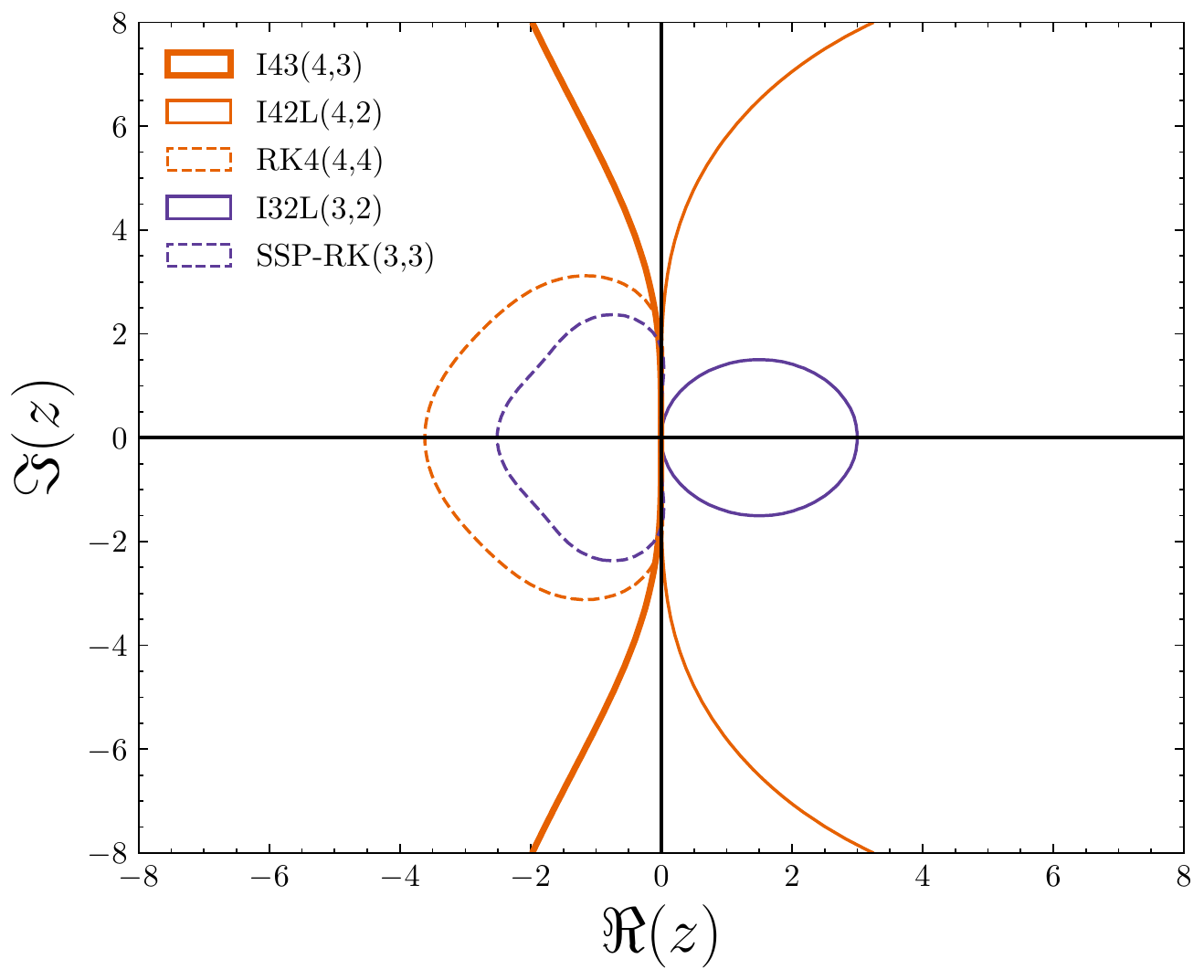}
	\caption{Stability region of the explicit \textbf{RK4(4,4)} and their respective implicit \emph{tableau} \textbf{I43(4,3)} and \textbf{I42L(4,2)}, and the explicit \textbf{SSP-RK(3,3)} with the implicit \emph{tableau} \textbf{I32L(3,2)}. The \textbf{I42L(4,2)} and \textbf{I32L(3,2)} are $L$-stable, while the \textbf{I43(4,3)} is only $L(\alpha)$-stable. The \textbf{I42L(4,2)} is the implicit \emph{tableau} that couples with the \textbf{RK4(4,4)} with the largest effective CFL that we have found.}
	\label{stability_functions}
\end{figure}

In the following subsections we present three IMEX schemes (which are summarized in Appendix \ref{A_summary_IMEX}) that satisfy these conditions and are constructed using the explicit part of the two most commonly used Runge-Kuttas: the standard fourth-order, \textbf{RK4(4,4)} \Mycite{originalRK4}, and the third-order Strong-Stability-Preserving scheme of Shu and Osher, \textbf{Shu-Osher SSP-RK(3,3)} \Mycite{shu_osher}. The stability region of these two explicit schemes, as well as their respective implicit \emph{tableau} are shown in Fig.~\ref{stability_functions}.
%
\subsubsection{\textbf{IMEX schemes coupling with the RK4(4,4)}}
First, we will try to find an implicit Butcher \emph{tableau} that couples with the explicit standard \textbf{RK4(4,4)} scheme \Mycite{originalRK4} in Table \ref{RK4}, which has a large $\text{CFL}=2$ and four stages such that $\text{CFL}_{\text{eff}} =1/2$. Again, our goal is to construct an IMEX RK that satisfies all the requirements commented in the previous subsection. This \textbf{RK4(4,4)}, in addition to being of high order and having a large $\text{CFL}_{\text{eff}}$, is close to be SSP (see, e.g., \Mycitep{gottlieb98}; \Mycitep{gottlieb03}).
\begin{table}[h]
	\caption{\emph{Tableau} for the explicit \textbf{RK4(4,4)} scheme. \label{RK4}}
	\begin{minipage}{1.4in}
		\begin{tabular} {c| c c c c}
			$0$ &  $0$ & $0$ & $0$ & $0$\\
			$1/2$ & $1/2$ & $0$ & $0$ & $0$ \\  
			$1/2$ & $0$& $1/2$ &$0$& $0$\\  
			$1$ & $0$ & $0$ & $1$ & $0$ \\  
			\hline \\[-1.0em] 
			& $1/6$& $1/3$ & $1/3$ & $1/6$ \\
		\end{tabular}
	\end{minipage}
\end{table}

The requirements described in the previous subsection impose strict constraints on the implicit RK. The fourth condition (cond. [\ref{c_equal_tilde_c}]) trivially implies that $c = \tilde{c} = (0,1/2,1/2,1)^T$. We restrict ourselves to stiffly accurate schemes such that $w_i=a_{4i}$ (for $i=1,\ldots,4$), while the condition $w=\tilde w=(1/6, 1/3, 1/3, 1/6)$ (cond. [\ref{w_equal_w}]) guarantees all mixed third order conditions (which are not shown here, but can be found in \Mycitep{pareschi05}). At this point, we can summarize the implicit scheme (i.e., the only unknown part) as the Butcher \emph{tableau} shown in Table \ref{general_tableau_RK4}.
\begin{center}
	\begin{table}[h!]
		\caption{General implicit \emph{tableau} that couples with the explicit \textbf{RK4(4,4)} scheme.} \label{general_tableau_RK4}
		\begin{minipage}{2.9in}
			\begin{tabular} {c|c c c c}
				$0$ &  $0$ & $0$ & $0$ & $0$\\
				$1/2$ & $a_{21}$ & $1/2-a_{21}$ & $0$ & $0$ \\  
				$1/2$ & $a_{31}$& $a_{32}$ &$1/2-a_{31}-a_{32}$& $0$\\  
				$1$ & $1/6$ & $1/3$ & $1/3$ & $1/6$ \\  
				\hline \\[-1.0em] 
				 & $1/6$ & $1/3$ & $1/3$ & $1/6$ \\  
			\end{tabular}
		\end{minipage}
	\end{table}
\end{center}
The assumption that $\hat{A}$ is invertible imposes two further constraints,
\begin{equation}\label{eq_cond_A_invertible}
	a_{31}+a_{32} \neq 1/2 ~~~ \text{and} ~~~ a_{21}\neq 1/2 \,.
\end{equation}
We can see that, after all these requirements, we are left with only three degrees of freedom, $\{a_{21}, a_{31}, a_{32}\}$, in the general case.
Since the explicit RK is fourth order, one of the third order conditions (eqs. \ref{eq_RKcond3_e}, \ref{eq_RKcond3}) of the implicit RK (i.e., $w^T\, c \,c = 1/3$ ) is also trivially satisfied. The other third order condition (i.e., $w^T A c = 1/6$), reduces to
\begin{equation}\label{eq_third_order_condition_implicit}
    a_{21} + a_{31} = 1/2 \,.
\end{equation}
Furthermore, by imposing $\hat{e}^T\hat{A}^{-1} a =0$ (eq. \ref{eq_additional_cond_RK_infinity}) to satisfy the condition (eq. \ref{eq_stability_function_RK_infinity}) for CK schemes, we obtain the following constraint
\begin{equation}\label{eq_L_stab_condition}
	a_{21}=\frac{6a_{31}+2a_{32}-1}{20a_{31}+20a_{32}-6} \,.
\end{equation}
Therefore, even without trying to maximize the CFL condition yet, only two degrees of freedom remain for second order IMEX schemes, and only one for third order IMEX schemes. 

Several options are possible for the choice of coefficients. The preferred case would be to both satisfy the third order condition (eqs. \ref{eq_RKcond3_e}, \ref{eq_RKcond3}, \ref{eq_third_order_condition_implicit}) and the stability condition (eq. \ref{eq_L_stab_condition}) simultaneously, leading to the existence condition,
\begin{equation}
	20a_{32}^{2} + 12a_{32} - 3 \geq 0 \,,
\end{equation}
yielding $a_{32} \in \left\{ \mathbb{R} - \left(-\sqrt{6}/5 -3/10, \sqrt{6}/5 +3/10 \right) \right\}$. Taking $a_{32} = 4/21$, we obtain a third order, $L(\alpha)$-stable, asymptotically preserving, implicit Butcher \emph{tableau} denoted as \textbf{I43(4,3)} in Table \ref{I43L}.
\begin{table}[t]
	\caption{\emph{Tableau} for the implicit \textbf{I43(4,3)} scheme. \label{I43L}}
	\begin{minipage}{1.4in}
		\begin{tabular} {c | c c c c}
			$0$ &  $0$ & $0$ & $0$ & $0$\\
			$1/2$ & $1/3$ & $1/6$ & $0$ & $0$ \\  
			$1/2$ & $1/6$& $4/21$ &$1/7$& $0$\\  
			$1$ & $1/6$ & $1/3$ & $1/3$ & $1/6$ \\  
			\hline \\[-1.0em] 
			 & $1/6$& $1/3$ & $1/3$ & $1/6$ \\
		\end{tabular}
	\end{minipage}
\end{table}

\begin{figure}[t!]
	\centering
	\includegraphics[width=0.48\textwidth]{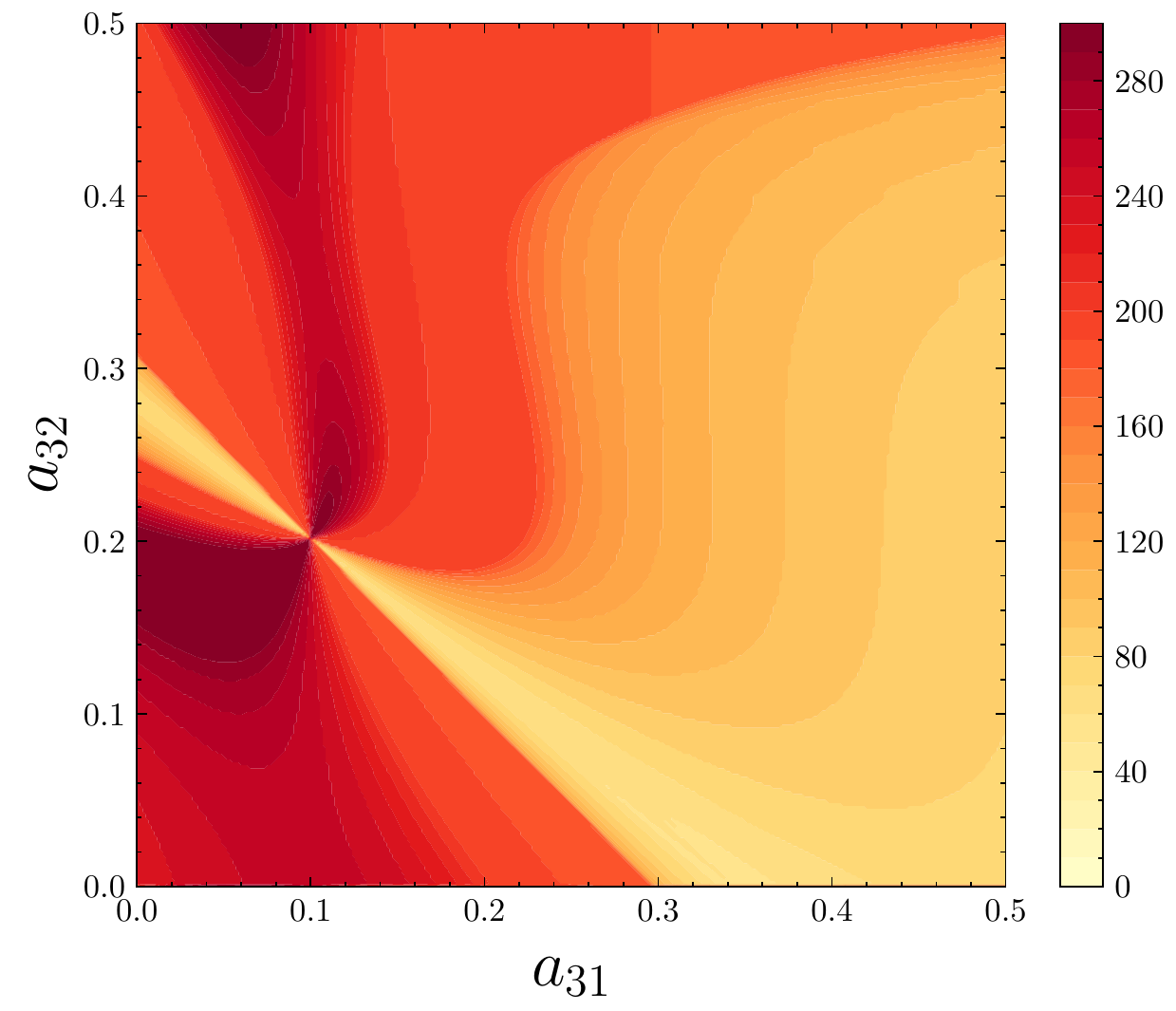}
	\caption{Stability regions within the domain $a_{31}, a_{32} \in \left[0, \frac{1}{2}\right] $ for second order $L$-stable IMEX schemes that couple with the explicit \textbf{RK4(4,4)} presented in Table \ref{RK4}. A random distribution of $300$ points lying inside the domain $\Im(z) \in [-8,8]$ and $\Re(z) \in [-4, 2]$ assigns a value $1$ or $0$ to those falling inside or outside respectively of the stability region (eq. \ref{eq_stability_function_RK}).}\label{region1}
\end{figure}

For second order schemes there is an additional degree of freedom. Therefore, our goal is to find which combination $\{ a_{31}, a_{32}\}$ maximizes the stability region  $\mathcal{R}=\mathcal{R}(z, a_{31}, a_{32})$. Our expectation is that such a choice will maximize the $\text{CFL}_\text{eff}$. Since the diagonal coefficients in the implicit matrix have to be positive (i.e., $a_{21}\leq1/2$ and $a_{31}+a_{32}\leq1/2$), we will only consider values of $a_{31}, a_{32} \in \left[0, \frac{1}{2}\right]$.
We have computed numerically the area of the regions that satisfy the stability condition (eq. \ref{eq_L_stab_condition}) for different values of these parameters. The value of this integral is displayed in Fig.~\ref{region1} as a function of the parameters $\{a_{31},a_{32}\}$. 
Among the options we explored, one of the simplest, which is $L$-stable, is the \textbf{I42L(4,2)}, represented in Table \ref{I42L}. 
\begin{table}[ht]
	\caption{\emph{Tableau} for the implicit \textbf{I42L(4,2)} scheme. \label{I42L}}
	\begin{minipage}{1.4in}
		\begin{tabular} {c | c c c c}
			$0$ &  $0$ & $0$ & $0$ & $0$\\
			$1/2$ & $1/4$ & $1/4$ & $0$ & $0$ \\  
			$1/2$ & $0$& $1/6$ &$1/3$& $0$\\  
			$1$ & $1/6$ & $1/3$ & $1/3$ & $1/6$ \\  
			\hline \\[-1.0em] 
			 & $1/6$& $1/3$ & $1/3$ & $1/6$ \\
		\end{tabular}
	\end{minipage}
\end{table}
The stability region of the two implicit \emph{tableau}, \textbf{I43(4,3)} and \textbf{I42L(4,2)}, as well as the one of the explicit \textbf{RK4(4,4)}, are displayed in Fig.~\ref{stability_functions}.

\subsubsection{\textbf{IMEX scheme coupling with the SSP-RK(3,3)}}

We are also interested on finding at least a second order implicit RK that couples with the explicit \textbf{Shu-Osher SSP-RK(3,3)} \Mycite{shu_osher} presented in Table \ref{RK3SSP}. It has been shown that this is the third-order SSP Runge-Kutta scheme with the largest $\text{CFL}=1$ and three stages such that $\text{CFL}_{\text{eff}} =1/3$. Notice that IMEX-SSP schemes, although not in the CK form, were already studied in \Mycitep{pareschi03}.
\begin{table}[h]
	\caption{\emph{Tableau} for the explicit \textbf{SSP-RK(3,3)} scheme. \label{RK3SSP}}
	\begin{minipage}{1.4in}
		\begin{tabular} {c | c c c }
			$0$ &  $0$ & $0$ & $0$ \\
			$1$ & $1$ & $0$ & $0$\\  
			$1/2$ & $1/4$ & $1/4$ & $0$  \\  
			\hline \\[-1.0em] 
			 & $1/6$& $1/6$ & $2/3$ \\
		\end{tabular}
	\end{minipage}
\end{table}

After imposing the trivial conditions $c=\tilde{c}=(0,1,1/2)^T$ (cond. [\ref{c_equal_tilde_c}]),  stiffly accurate schemes such that $w_i=a_{3i}$ (for $i=1,\ldots,3$) , and  $w=\tilde w=(1/6, 1/6, 2/3)$ (cond. [\ref{w_equal_w}]), the most general Butcher \emph{tableau} is given by Table \ref{general_tableau_RK3}.
\begin{table}[h]
	\caption{General implicit \emph{tableau} that  couples with the explicit \textbf{SSP-RK(3,3)} scheme. \label{general_tableau_RK3}}
	\begin{minipage}{1.4in}
		\begin{tabular} {c | c c c }
		$0$ &  $0$ & $0$ & $0$ \\
		$1$ & $a_{21}$ & $1-a_{21}$ & $0$ \\  
		$1/2$ & $1/6$ & $1/6$ & $2/3$ \\  
		\hline \\[-1.0em] 
		& $1/6$ & $1/6$ & $2/3$ \\  
		\end{tabular}
	\end{minipage}
\end{table}

In this case there is only one degree of freedom for the coefficient $a_{21}$. The second order conditions are trivially satisfied, while the stability condition requires again 
$\hat{e}^{T}\hat{A}^{-1}a=0$ (eq. \ref{eq_additional_cond_RK_infinity}) with $\hat{e}=(0,1)^{T}$, leading to the unique condition $a_{21} = 1/2$.

Unfortunately, the IMEX with $a_{21}=1/2$ is only second order, thus we are able to conclude that there will only be one implicit Butcher \emph{tableau} that couples with the \textbf{SSP-RK(3,3)} and meets all our requirements. Therefore, we obtain the following second order $L$-stable implicit scheme called \textbf{I32L(3,2)} in Table \ref{I32L}. Finally, the stability regions of the implicit \textbf{I32L(3,2)} and the explicit \textbf{SSP-RK(3,3)} are shown in Fig.~ \ref{stability_functions}.
\begin{table}[h]
	\caption{\emph{Tableau} for the implicit \textbf{I32L(3,2)} scheme. \label{I32L}}
	\begin{minipage}{1.4in}
		\begin{tabular} {c | c c c }
			$0$ &  $0$ & $0$ & $0$ \\
			$1$ & $1/2$ & $1/2$ & $0$ \\  
			$1/2$ & $1/6$ & $1/6$ & $2/3$\\  
			\hline \\[-1.0em] 
			& $1/6$& $1/6$ & $2/3$\\
		\end{tabular}
	\end{minipage}
\end{table}
%
%
%
\subsection{Spatial Discretization Scheme}\label{spatial_discretization_scheme}
In our simulations, the spacetime, the fluid and the neutrinos will be integrated in time by using the same generic IMEX RK described before in Table \ref{I42L} (for the complete IMEX scheme see Table \ref{A_final_IMEX4}). However, the MoL allow us to use the space discretization for each field which is most efficient for each case. 

We use fourth order-accurate operators for the spatial derivatives of the Einstein equations,  supplemented with sixth-order Kreiss-Oliger dissipation (more details in \Mycitep{sim18}; \Mycitep{daniele19b}). The fluid equations can be written formally in conservation law form, namely 
\begin{eqnarray}\label{eq_PDEequationdecomposed}
	\partial_t \textbf{U} + \partial_k \textbf{F}^{k}(\bf{U}) = \textbf{S} (\bf{U}) \,, 
\end{eqnarray}
where ${\bf U}$ is the list of evolved fields. Here,  $\textbf{F}^{k}(\bf{U})$ and $\textbf{S}({\bf U})$ are their corresponding fluxes and sources, which might be non-linear but depend only on the fields and not on their derivatives. These equations are discretized with a high-resolution shock-capturing (HRSC) method based on the Lax-Friedrichs flux splitting formula \Mycite{shu98} and the fifth-order reconstruction method MP5 \Mycite{mp5}.
	
The M1 equations have the same structure, apart from the stiff terms, as the fluid equations. They can also be written as a system of balance laws, and HRSC are needed to deal with possible shocks.
In addition, as discussed in \Mycite{rad2022}, the choice of a spatial discretization scheme for these equations (eqs. $\ref{eq_evolE}$, $\ref{eq_evolF}$) is constrained by the \emph{ill-posedness} of the diffusion equation (\Mycitep{hiscock85}; \Mycitep{rider02}; \Mycitep{andersson11}), which requires that the fluxes reduce to a consistent second-order finite difference scheme in the optically thick limit. The fluxes are constructed following the \emph{flux-splitting} approach described in LeVeque's seminal book \Mycite{leveque}, i.e., as a linear combination of a non-diffusive high order (second order) flux $F^{\text{HO}}$ that works well in smooth regions, and a diffusive low order Lax-Friedrichs (first order) flux $F^{\text{LO}}$ that behaves well near discontinuities. This robust and easy to implement spatial integration scheme has also been recently used in \Mycite{rad2022}. 

The partial derivatives in the balance law (eq. \ref{eq_PDEequationdecomposed}), at the grid location $x_i$, can be written directly as a conservative finite-difference approximation. Considering for simplicity just a single field, with fluxes only along the x-direction, it reduces to:
\begin{equation}
	\partial_t U_i = - \frac{F_{i + 1/2} - F_{i-1/2}}{\Delta x} + S_i \,,
\end{equation}
where $F_{i + 1/2}$ and $F_{i - 1/2}$ are numerical fluxes computed at the interfaces of $x_i$, i.e., $x_i \pm \frac{\Delta x}{2}$, respectively. Additionally, we define a flux limiter, $\Phi (\textbf{U})$, that switches between the high and low order flux and a coefficient, $A_{ i + 1 / 2} (\Delta x_i , \kappa_a , \kappa_s )$, to switch off the diffusive correction at high optical depth. The simplest finite volume (FV) reconstruction that is asymptotically preserving in the limit of the diffusion equation (i.e.,  $F_{i+1/2} \approx F^{\text{HO}}_{i+1/2}$ in the optically thick limit) is given by
\begin{equation}
	\hat{F}_{i+1/2} = F^{\text{HO}}_{i+1/2} - A_{i+1/2}\left[1 - \Phi_{i+1/2}\right] \left(F^{\text{HO}}_{i+1/2} - F^{\text{LO}}_{i+1/2}\right) \,,
\end{equation}
where
\begin{eqnarray} 
	F^{\text{HO}}_{i+1/2} &=& \frac{1}{2} \left[f(u_i) + f(u_{i+1}) \right] \,,\\ 
	F^{\text{LO}}_{i+1/2} &=& \frac{1}{2} \left[f(u_i) + f(u_{i+1}) \right] - \frac{\lambda_{i+1/2}}{2}\left[u_i + u_{i+1}\right] \,, \nonumber \\
	\Phi_{i+1/2} &=& \text{max}\left[0, \text{min}\left(1, 2 \frac{u_i - u_{i-1}}{u_{i+1}-u_i}, \nonumber 2 \frac{u_{i+2}-u_{i+1}}{u_{i+1}-u_i}  \right)   \right] \,, \\ \nonumber
	A_{i+1/2} &=& \text{tanh}\left[\frac{1}{\kappa_{i+1/2} \Delta x}\right] \,. \nonumber
\end{eqnarray}
This average coefficient, $\kappa_{i+1/2}$, can be computed as
\begin{equation}
	\kappa_{i+1/2} = \frac{1}{2} \left[  (\kappa_{a})_i + (\kappa_{a})_{i+1} + (\kappa_{s})_i + (\kappa_{s})_{i+1} \right] \,.
\end{equation}
In some cases, there might appear odd-even oscillations due to a decoupling of consecutive grid points in the spatial discretization scheme. A necessary and sufficient condition for this problem to appear is that \Mycite{odd_even}
\begin{equation}
	g_i ~ g_{i-1} < 0   ~~~\text{and}~~~
	g_i ~ g_{i+1} < 0   \,,
\end{equation}
where $g_i \equiv u_{i+1} + u_{i-1} - 2 u_{i}$ is the concaveness of the evolved field $u$ at the grid point $x_i$. We note that a similar condition can be found in \Mycite{rad2022}. When the above conditions are satisfied simultaneously, we set $A_i=1$, which is enough to cure the problem. 
	
The characteristic velocities, $\lambda_{i+1/2}$,  can be calculated both in the thin and thick regimes \Mycite{shi11}, and then interpolate using the Minerbo closure
\begin{equation}\label{minerbo_velocity}
	\lambda_i = \frac{3 \chi - 1}{2} \lambda^{\text{thin}}_{i}
	+ \frac{3 (1- \chi)}{2} \lambda^{\text{thick}}_{i} \,,
\end{equation}
where
%
	\begin{eqnarray}
		\lambda^{\text{thin}}_{i} &=& \max \left( |\beta_i| \pm \alpha \frac{|F_i|}{\sqrt{F_k F^k}} , \, |\beta_i| + \alpha E \frac{ |F_i|}{{F_k F^k}}
		\right) ~\,, \nonumber\\
		\lambda^{\text{thick}}_{i} &=& |\beta_i| \nonumber +  \frac{2 W^2 |v_i| \pm \sqrt{(2 W^2 + 1) \alpha \gamma^{ii} - 2 W^2 v_i v^i}}{2 W^2 + 1} ~~ \,.\nonumber	 
	\end{eqnarray}
%
In the numerical simulations presented here we have not found any significant difference between using the full analytical expressions for the characteristic velocities versus using the maximum value of the speed of light between neighboring cells. Nevertheless, since this might change in more complicated scenarios, we have implemented the full expressions.
%
\subsection{Implicit Treatment of Stiff Source Terms: Radiation-Matter Coupling}\label{SS_newton_raphson}
As discussed in subsec. \ref{SS_M1_evol_eqs}, we have split the vector of fields, $(\bf U)$, into a stiff, $(\bf V)$, and a non-stiff, $(\bf W)$, part. The evolution procedure to compute each step $U^{(i)}$ can be divided in two substeps:
\begin{enumerate}
	\item Compute the explicit intermediate values $\{\bf{V^*},\bf{W^*}\}$, i.e.,
	\begin{eqnarray}\label{eq_first_step}
		{\bf W}^{*} = {\bf W}^n &+& \Delta t~ \sum_{j=1}^{i-1}~ {\tilde{a}}_{ij} F_W({\bf U}^{(j)}) \,,
		\nonumber \\
		{\bf V}^{*} = {\bf V}^n &+& \Delta t~ \sum_{j=1}^{i-1}~ {\tilde{a}}_{ij} F_V({\bf U}^{(j)}) 
		\nonumber \\
		&+& \Delta t~ \sum_{j=1}^{i-1}~ a_{ij} \frac{1}{\epsilon^{(j)}} R_V({\bf U}^{(j)}) \,,
	\end{eqnarray} 
	where we have defined $ \epsilon^{(j)} = \epsilon({\bf W}^{(j)})$.
	\item Compute the implicit part, involving only ${\bf V}$, by solving the implicit equation
	\begin{eqnarray}
		{\bf W}^{(i)} &=& {\bf W}^{*} \,,\\
		{\bf V}^{(i)} &=& {\bf V^*} 
		+ a_{ii}~\frac{\Delta t}{\epsilon^{(i)}}~R_V({\bf V}^{(i)},{\bf W}^{(i)}) \,.
		\label{eq_second_step}
	\end{eqnarray}
	The equation (eq. \ref{eq_second_step}) can be solved in different ways depending on the nature of the source terms. The simplest case is when  they depend linearly on the evolution fields, such that the stiff part can be written in the following way,
	\begin{eqnarray}\label{eq_stiff_part}
		R_V({\bf V},{\bf W}) = \mathbb{A} ({\bf W}) {\bf V} + S(\bf{W}) \,.
	\end{eqnarray}  
	Then, the implicit equation can be solved just by inverting a matrix, namely
	\begin{eqnarray}\label{eq_invert_matrix_lin}
		{\bf V}^{(i)} &=& \mathbb{M} ~
		\left( {\bf V^*} + a_{ii}~\frac{\Delta t}{\epsilon^{(i)}}~S({\bf W^*}) \right) \,,
		\\
   \mathbb{M} &\equiv& \left[I - a_{ii}~\frac{\Delta t} {\epsilon^{(i)}} \mathbb{A}({\bf W^*})\right]^{-1} \,.
	\end{eqnarray}
    If the source terms depend non-linearly on the evolution fields, the generic way to solve it is by finding the zeros of the system of equations
	\begin{eqnarray}\label{eq_nonlinearstiff}
	{\bf G} &=& {\bf V}^{(i)} - {\bf V^*} 
	- a_{ii}~\frac{\Delta t}{\epsilon^{(i)}}~R_V({\bf V}^{(i)},{\bf W}^{(i)}) = 0 \,.
	\end{eqnarray}
	A multidimensional Newton-Raphson solver, involving the inverse of the Jacobian of ${\bf G}$ with respect to ${\bf V}$, can be employed to find such solutions. In particular, the solution at the iteration $n$ can be calculated from the solution at the iteration $n-1$ as follows
	\begin{equation}
	{\bf V}_{n+1}^{(i)}  = {\bf V}_n^{(i)} - \left(\frac{\partial {\bf G}  }{\partial {\bf V}}\right)_n^{-1} {\bf G}_n \,,
	\end{equation}
	and where the Jacobian of ${\bf G}$ can be written in terms of the Jacobian of $R_V$ easily, namely
	\begin{equation}
	\left(\frac{\partial {\bf G}  }{\partial {\bf V}}\right) = {\bf I}  - a_{ii}~\frac{\Delta t}{\epsilon^{(i)}} \left(\frac{\partial {R_V}  }{\partial {\bf V}}\right) \,.
	\end{equation}
	\item Once the solution ${\bf V}^{(i)}$ is found, calculate the stiff term $R_V({\bf V}^{(i)},{\bf W}^{(i)})$ for the next RK sub-step.
\end{enumerate}

In the specific case of the $\text{M}1+\text{N}0$ equations, the fields and the stiff terms are given by
\begin{equation}
\tmmathbf{V}= \left(\begin{array}{c}
\sqrt{\gamma} n  \Gamma\\
\sqrt{\gamma} E\\
\sqrt{\gamma} F^i
\end{array}\right) \,, \qquad 
R_V = \left(\begin{array}{c}
\alpha {\cal S}_N\\
\alpha {\cal S}_n\\
\alpha {\cal S}^i
\end{array}\right) \,, \\
\end{equation}
where
\begin{eqnarray}
&& {\cal S}_N=\eta^0 - \kappa_a^0 n \,, \\\label{stiff_terms}
&&{\cal S}_n = - n_a {\cal S}^a 
 = W \left[ (\eta + \kappa_s J) - (\kappa_a + \kappa_s) (E - F_i v^i) \right] \,, \nonumber \\
&&{\cal S}^i = \gamma^i_b {\cal S}^b = W (\eta - \kappa_a J) v^i - (\kappa_a + \kappa_s) H^i \,.
\nonumber
\end{eqnarray}
Clearly, the stiff terms for the field $n \Gamma$ are linear and  independent of the others, so they can be solved easily following (eqs. \ref{eq_invert_matrix_lin}). The other stiff terms are coupled and non-linear, so the generic solution, by using a multidimensional Newton-Raphson algorithm,  requires the calculation of the Jacobian ${\cal J} = \frac{\partial R_V}{\partial {\bf V}^{(i)}}$, which can be found explicitly in \Mycite{rad2022} (reprinted in Appendix \ref{A_jacobian_terms}).
%
\section{Numerical Tests}\label{S_numerical_tests}
To validate the accuracy of our computational and numerical scheme we have performed a set of demanding tests to asses our implementation of the equations. Among the many possible tests found in the literature (see, e.g., \Mycitep{smit97}; \Mycitep{audit02}; \Mycitep{rampp02}; \Mycitep{vaytet11}; \Mycitep{sado13}; \Mycitep{rad13}; \Mycitep{mck14}; \Mycitep{nagakura14}; \Mycitep{fou15}; \Mycitep{kuroda16}; \Mycitep{skinner19}; \Mycitep{weih20}; \Mycitep{chan20}; \Mycitep{anninos20}; \Mycitep{rad2022}), here we have included only some of the most representative for our purposes. Essentially, we have followed the recent work \Mycite{rad2022} which has set a new standard by being among the first (as well as \Mycitep{anninos20}) to include a full treatment of the nonlinear terms in the radiation-matter coupling (see Appendix \ref{A_jacobian_terms}). These numerical tests have been performed using \texttt{MHDuet}, our  (publicly available) code for evolving a spacetime coupled to matter (i.e., in this case, a magnetized perfect fluid and neutrinos). \texttt{MHDuet} is generated by the open-source platform \texttt{SIMFLOWNY} (\Mycitep{sim1}; \Mycitep{sim2}; \Mycitep{sim3}) and runs under the \texttt{SAMRAI} infrastructure (\Mycitep{samrai1}; \Mycitep{samrai2}), which provides parallelization and FMR/AMR. The code has been extensively tested for different MHD and strong-gravity scenarios (\Mycitep{sim18}; \Mycitep{daniele19b}; \Mycitep{daniele20}; \Mycitep{steve20}; \Mycitep{leakage22}), so here both the spacetime and fluid are set to trivial solutions and we will focus only on the neutrino evolution.
 \begin{figure}[t!]
	\centering
	\includegraphics[width=\columnwidth]{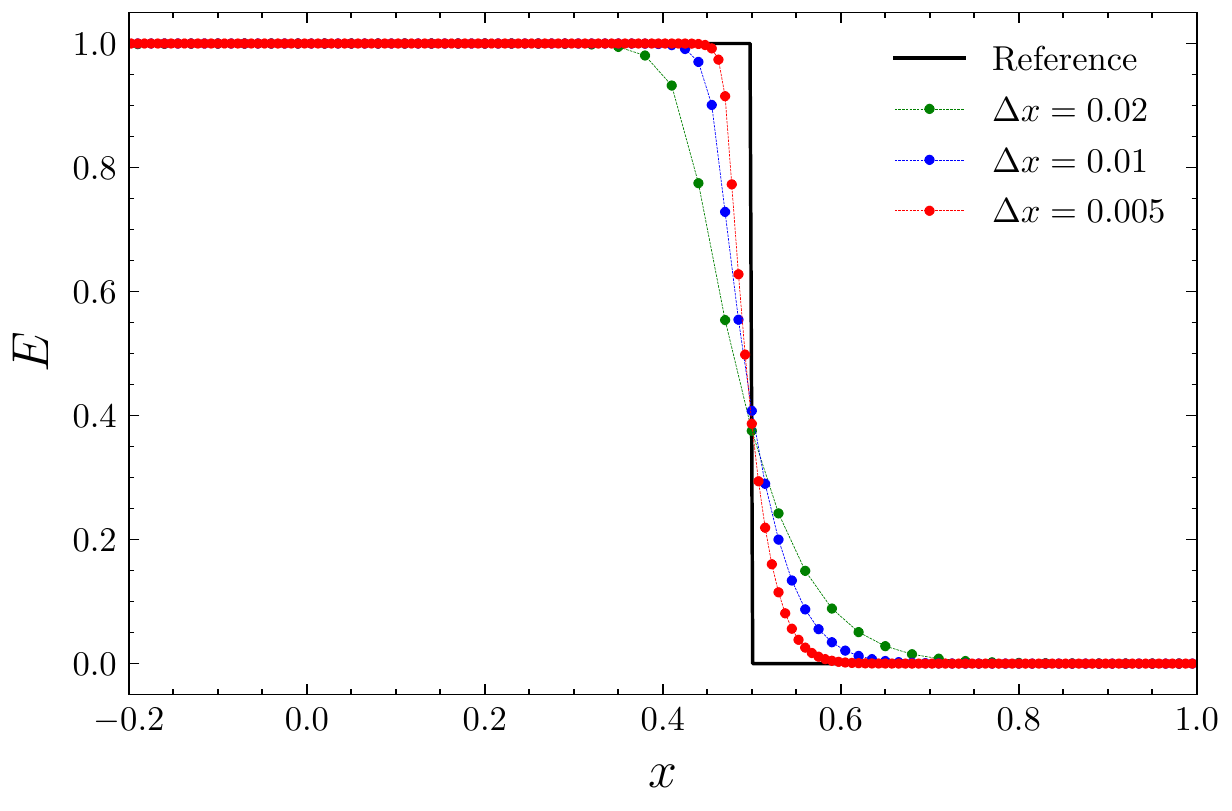}
	\caption{{\em Test 1. Optically Thin Advection Through a Velocity Jump}. A step-like radiation density profile is advected through a large	velocity discontinuity, with a relative Lorentz factor of $W=7$, between the two parts of the domain ($x < 0$ and $x>0$). The numerical solution is compared for different resolutions with the exact reference solution at the time $t = 1$.}
	\label{T1_final}
\end{figure}
%
\subsection*{Test 1. Optically Thin Advection Through a Velocity Jump} \label{T1}
This first test deals with the propagation of a radiation beam in an optically thin medium (\Mycitep{rad2022}). We consider a Heaviside function, $\text{H}(u)$, to construct a step profile as initial data, namely
\begin{eqnarray}
	&E&(t=0, \mathbf{x}) = F^{x}(t=0, \mathbf{x}) = \text{H}(\mathbf{x} + 1/2) \,.
\end{eqnarray}
We will consider the propagation of such a solution in the optically thin limit,
 \[
	 	\kappa_{s}(\mathbf{x})=\kappa_{a}(\mathbf{x})=\eta(\mathbf{x})=0 \,.
 \]
Here, the main numerical challenge is that the medium is moving with a very high Lorentz factor, i.e., the velocity of the background fluid is given by
\[
 	v^{x}(\mathbf{x}) =
 	\begin{cases}
 		~~0.87, ~~~  x < 0 \\
 		-0.87, ~~~ x > 0
 	\end{cases}
\]
 which implies a relative Lorentz factor of $W=7$ between the two parts of the domain ($x < 0$ and $x>0$), and a Lorentz factor $W=2$ in the mesh frame. We have employed a Cartesian domain with $x \in [-3,3]$, covered with different resolutions corresponding to $N = \{1600, 800, 400, 200, 100, 50 \}$ grid points with a $\text{CFL} =2$.
 
In Fig.~\ref{T1_final} we can see the radiation energy density profile at time $t=1$, after the beam has propagated through the velocity jump at $x=0$. From this figure it is clear that our numerical scheme is able to solve a discontinuous initial profile in a moving medium without creating artificial oscillations. As expected, due to the spatial discontinuity present in the field, the solution converges only at first order. 
%
\subsection*{Test 2. Diffusion Limit in a Scattering Medium}\label{T2}
One of the most important requirements that our numerical scheme needs to satisfy for successfully modeling BNS mergers is to be able to deal with the neutrino diffusion of the remnant in the post-merger phase. In this test (see, e.g., \Mycitep{pons00}; \Mycitep{audit02}; \Mycitep{rad13}; \Mycitep{mck14}; \Mycitep{kuroda16}; \Mycitep{weih20}; \Mycitep{rad2022}), we focus on how to deal with the diffusion of radiation when the opacity is high (optically thick limit) and the mean free path is small compared to the grid spacing $\Delta x$. This test is crucial to check the performance and expected accuracy of our IMEX scheme (see Table \ref{A_final_IMEX4}), since the profiles are initially smooth and no shocks are produced during the evolution that could degrade the accuracy.

Here, we consider the diffusion of a radiation beam that starts as a step function in a purely scattering, static medium. The initial data is given by:
\begin{align}
	&E(t=0, ~\mathbf{x}) = \text{H}(\mathbf{x} + 1/2) - \text{H}(\mathbf{x} - 1/2) \,, \\
	&F^i(t=0, \mathbf{x}) = v^i(t=0, ~\mathbf{x}) = 0 \,, \\
	&\kappa_{s}(\mathbf{x})=10^3, ~~~ \kappa_{a}(\mathbf{x}) = \eta(\mathbf{x})=0 \,.
\end{align}
We consider a domain $x \in [-2,2]$ resolved by different resolutions, corresponding to $N  = \{1600, 800, 400, 200, 100, 50\}$ grid points with a $\text{CFL} = 2$.

For timescales longer than the equilibrium time, the evolution of $E$ can be well approximated by the diffusion equation. The exact solution of which is given by
\begin{equation}\label{eq_T2_solution}
	E(t, \mathbf{x}) = \frac{1}{2} \left[ \text{erf}\left(\frac{\frac{1}{2} + \mathbf{x}}{\sqrt{4\tau t}}\right) - \text{erf}\left(\frac{\mathbf{x} - \frac{1}{2}}{\sqrt{4\tau t}}\right) \right] \,,
\end{equation}
where $\tau = \frac{1}{3\kappa_s}$ is the diffusion timescale. 
\begin{figure}[t!]
	\includegraphics[width=\columnwidth]{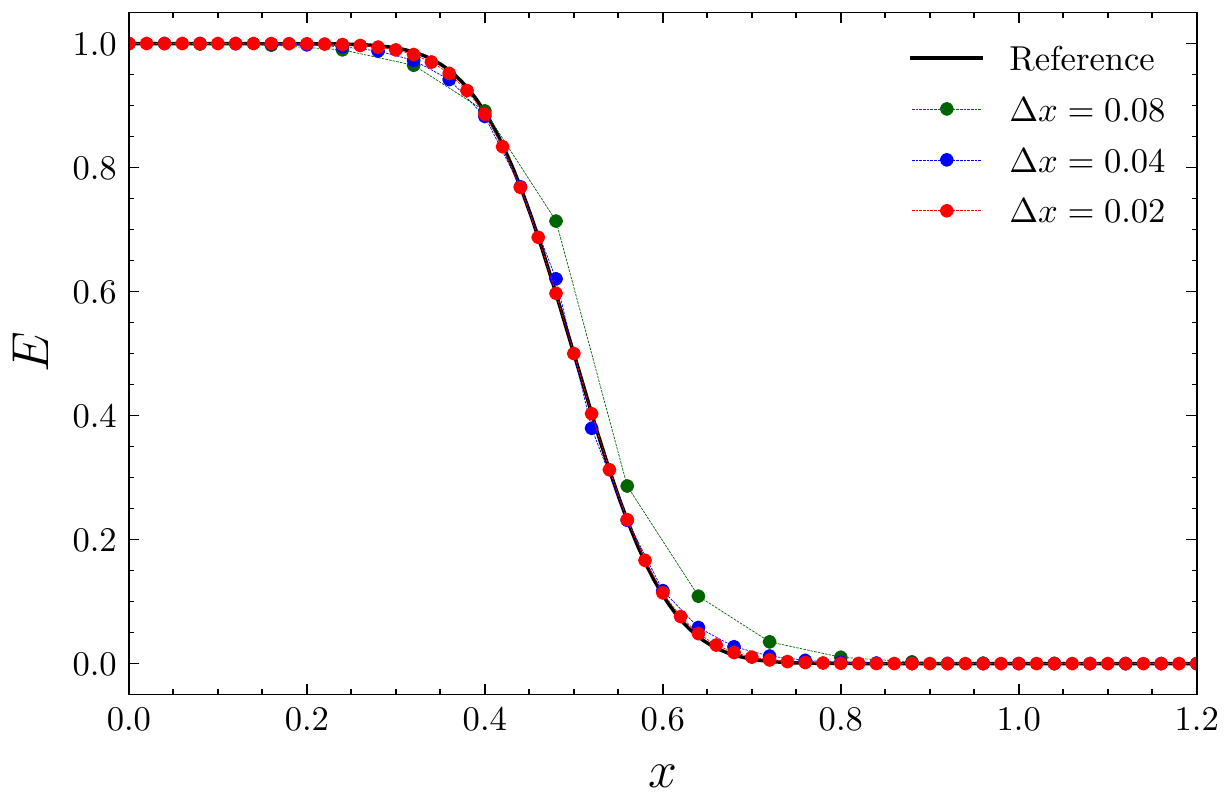}
	\caption{{\em Test 2. Diffusion Limit in a Scattering Medium}. The energy density is initialized with a gate function and slowly diffuses in a purely scattering medium with $\kappa_{s} = 10^3$. The numerical solution at time $t=10$ is compared again with different resolutions and with the exact reference solution given by (eq. \ref{eq_T2_solution}).}
	\label{T2_final}
\end{figure}

Fig.~\ref{T2_final} shows the profile of the radiation energy density at time $t=10$ for low and intermediate resolutions, compared to the semi-analytical solution of (eq. \ref{eq_T2_solution}). We can observe that as we increase the resolution the numerical solution becomes indistinguishable from the exact reference.
\begin{figure}[t!]
	\includegraphics[width=\columnwidth]{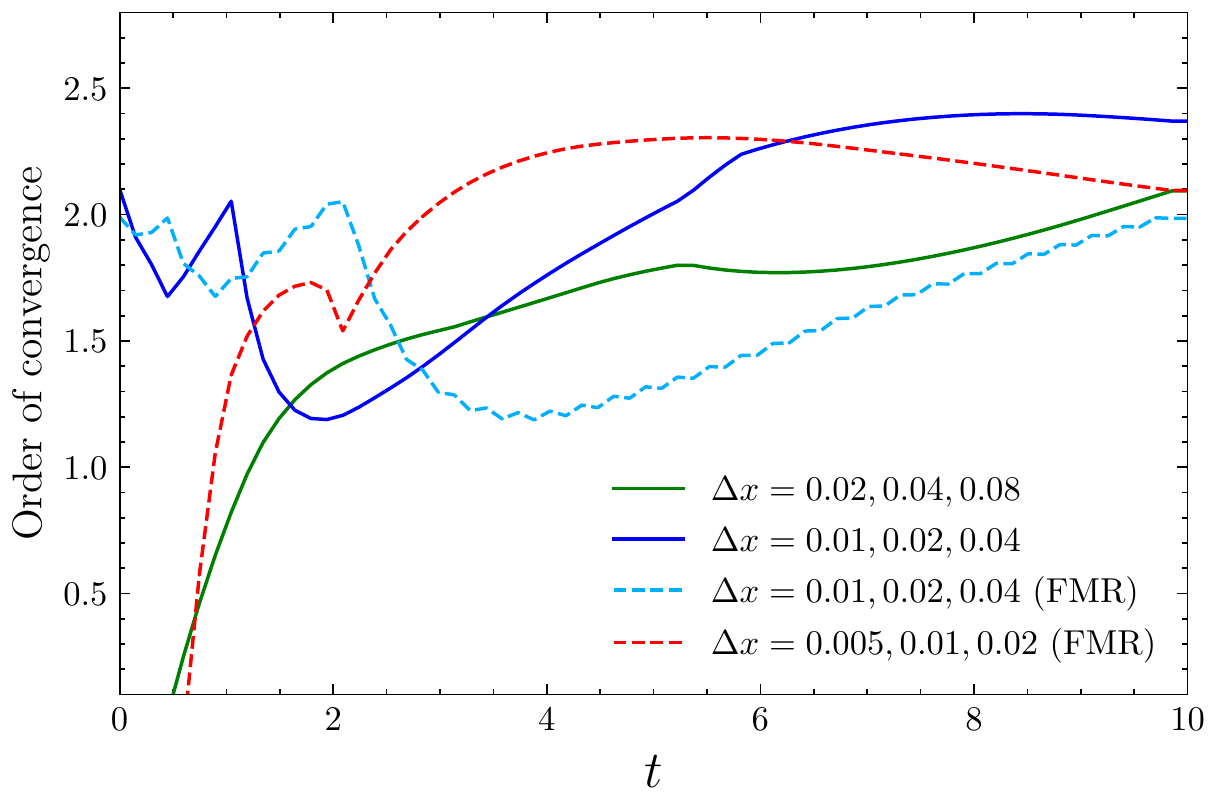}
	\caption{{\em Test 2. Diffusion Limit in a Scattering Medium}. Numerical convergence  for several resolutions without and with an extra FMR level, with twice the resolution of the coarsest level, placed at $x \in [-1,1]$. The solution shows approximately the expected second order convergence.}
	\label{convT2}
\end{figure}	
The numerical convergence shown in Fig. \ref{convT2} is approximately second order. This test shows quite thoroughly the behavior of our IMEX since the implicit stiff source terms dominate the right-hand sides. As a novelty in these tests and to check the implementation of IMEX schemes in \texttt{MHDuet} we perform several simulations with an additional FMR grid level, with twice the resolution of the coarsest level, placed at $x \in [-1,1]$. In Fig. \ref{convT2} we can observe that the second order of convergence is still preserved when using FMR. This result is very significant, since FMR/AMR are employed routinely in BNS merger simulations.
%
\subsection*{Test 3. Diffusion in a Moving Medium}\label{T3}
This test (see, e.g., \Mycitep{nagakura14}; \Mycitep{chan20}; \Mycitep{rad2022}) is probably the most demanding of the ones presented here, as it contains all the ingredients that have made this problem the focus of many authors for years. The test consists on a radiation energy density, with a Gaussian profile, which propagates in a purely scattering medium that is moving to the right with a relativistic velocity, namely
\begin{eqnarray}
	&E(t=0, \mathbf{x}) = e^{-9x^2} \,, ~~~
	v^x(t=0, \mathbf{x}) = 0.5 \,,\\
	&~~~\kappa_{s}(\mathbf{x}) =10^3 \,, ~~~~~~ \kappa_{a}(\mathbf{x}) = \eta(\mathbf{x})=0 \,.			
\end{eqnarray}		
The radiation fluxes, $F_{i}$, are initialized under the assumption that the radiation is fully trapped (i.e., $H^{a} = 0$). Using (eqs. \ref{eq_Jthick}, \ref{eq_Hithick}), we obtain the following relations
\begin{equation}
	J = \frac{3E}{4W^2 - 1} \,, \quad~~~~~ F_i = \frac{4}{3}JW^2v_i \,.
\end{equation}
In order to check the convergence of the numerical solutions we have considered different resolutions, corresponding to $N  = \{3200, 1600, 800, 400, 200, 100 \}$ grid points, covering a Cartesian domain $x \in [-4,4]$ with a $\text{CFL}=2$.
\begin{figure}[t!]
	\includegraphics[width=\columnwidth]{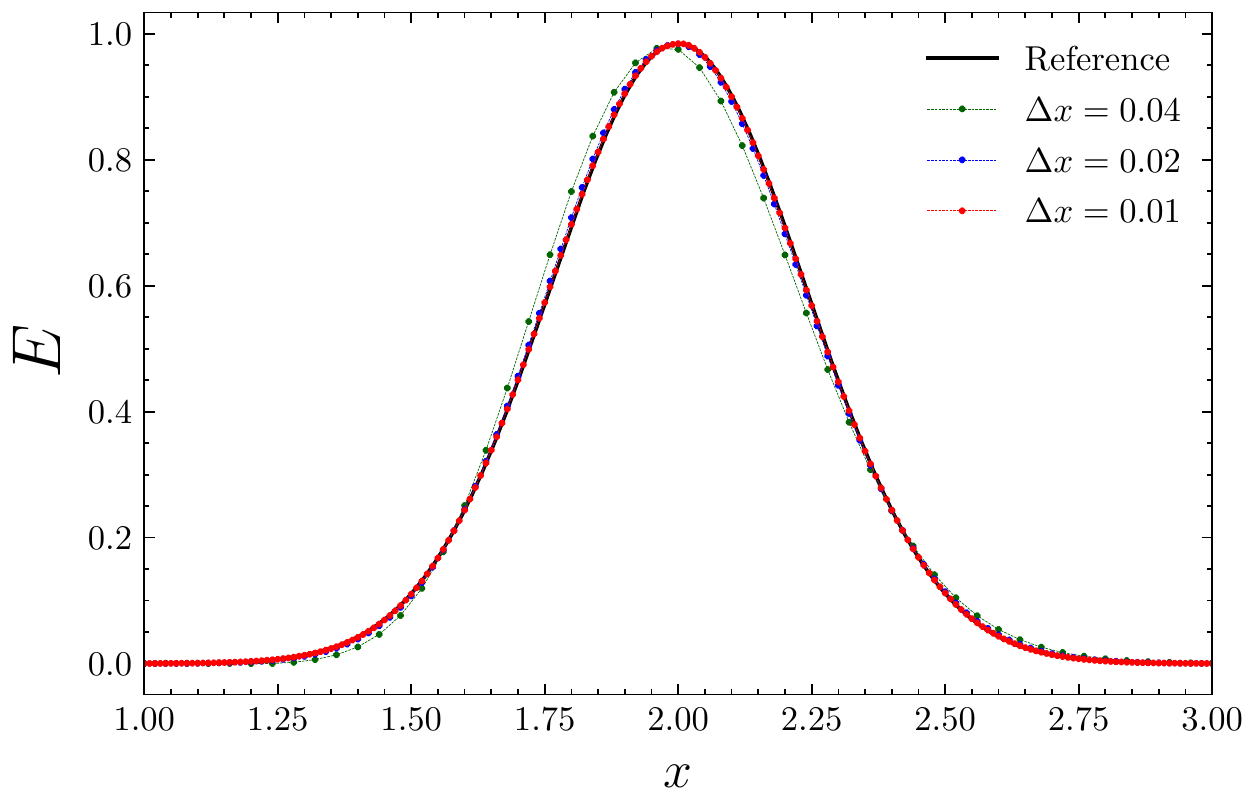}
	\caption{{\em Test 3. Diffusion in a Moving Medium}. The energy density is initialized as a Gaussian pulse of radiation which is advected while slowly diffuses in a purely scattering moving medium with $\kappa_{s} = 10^3$ and $v^x=0.5$. The profile at time $t=4$ is presented for three different resolutions and compared with the semi-analytic solution.}
	\label{T3_final}
\end{figure}

Fig. \ref{T3_final} shows the profile of the radiation energy density at time $t=4$ for different resolutions compared with the semi-analytic solution as in \Mycite{rad2022}. 
\begin{figure}[t!]
	\includegraphics[width=\columnwidth]{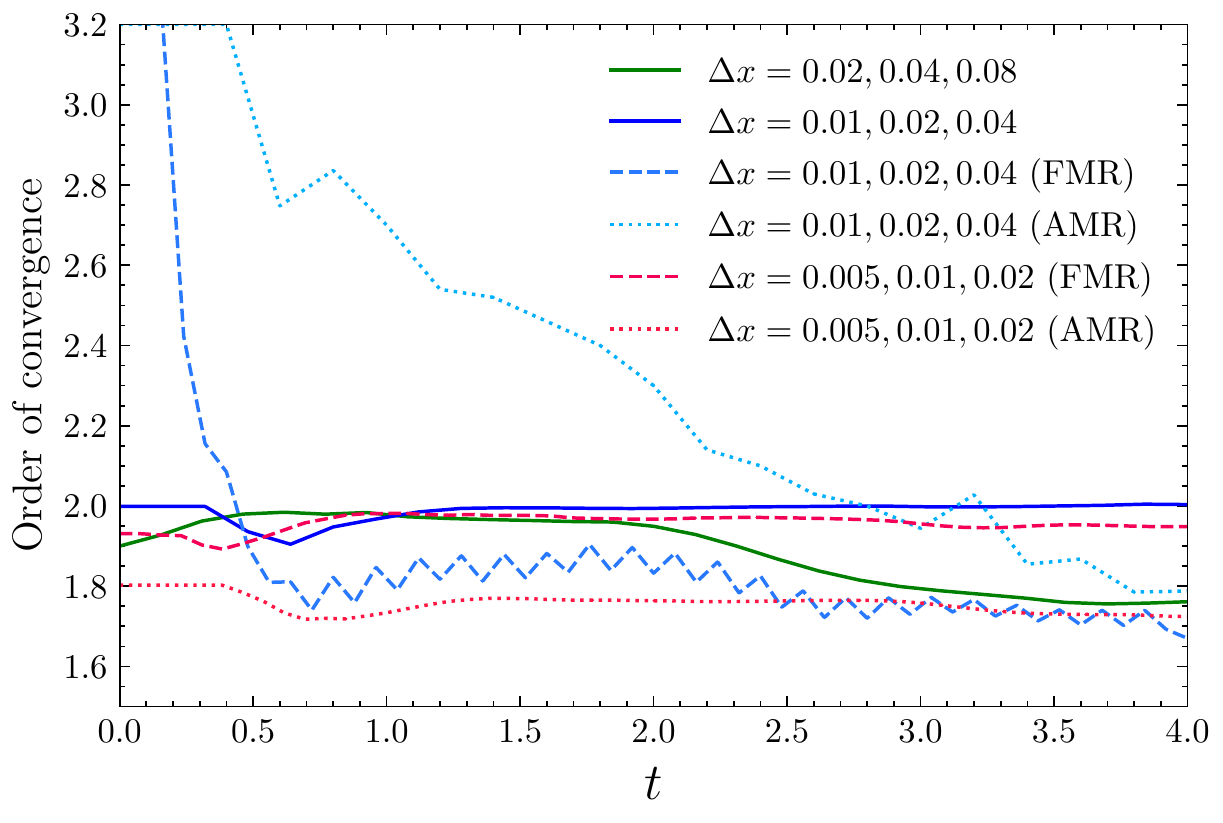}
	\caption{{\em Test 3. Diffusion in a Moving Medium}. Numerical convergence for several resolutions. We have also considered the case of: (i) an extra FMR level, with twice the resolution of the coarsest level, placed at $x \in [-0.75,1.25]$, and (ii) an AMR extra level following the pulse. We obtain the expected convergence order, between $1.8-2$, for all cases.}
	\label{T3_num_conv}
\end{figure}
Despite the pitfalls of this test, the results of our simulations show an approximately second-order convergence for several resolutions, as it is displayed in Fig.~\ref{T3_num_conv}. Furthermore, we have investigated the effect of an additional AMR grid level following the gaussian profile, observing still second order convergence. Finally, we explored the accuracy of the numerical solution as the Gaussian profile crosses an additional FMR level with twice the resolution of the coarsest level, placed at $x \in [0,75,1,25]$. As it was expected, the Gaussian profile is unaffected as it traverses the resolution change, still achieving second order convergence.
%
\subsection*{Test 4. Shadow Test}\label{T4}
This two-dimensional test (see, e.g., \Mycitep{sado13}; \Mycitep{mck14}; \Mycitep{just15}; \Mycitep{kuroda16}; \Mycitep{anninos20}; \Mycitep{weih20}; \Mycitep{rad2022}) shows the advantage of the M1 scheme versus a less accurate flux-limited diffusion scheme (FLD), namely the ability of an opaque object to generate a shadow when being illuminated by radiation. We set up a beam of radiation coming from the left boundary, i.e.,
\begin{equation}
	E(t=0,x_{\text{left}},y) =  F^{x}(t=0,x_{\text{left}},y) = 1 \,.
\end{equation} 		
and zero elsewhere. This beam interacts with a semi-transparent cylinder of radius, $R_{\star}=1$, centered at the origin. Inside the cylinder, the absorption opacity is $\kappa_{a}=1$. Absorption is zero elsewhere, as well as the scattering opacity and the emissivity, i.e.,
\begin{eqnarray}
	&\kappa_{a}(\mathbf{x}, \mathbf{y}) =& 			
	\begin{cases} 
		1, &  \sqrt{x^2 + y^2} < R_{\star},\\
		0, & \sqrt{x^2 + y^2} > R_{\star}.
	\end{cases} \\
	&\kappa_{s}(\mathbf{x},\mathbf{y}) =& \eta(\mathbf{x},\mathbf{y}) = 0 
\end{eqnarray}

 The simulation is performed in a Cartesian domain with $ (x,y) \in [-2,4] \times [-3,3]$ and resolved by different resolutions, corresponding to $ \{480^2, 240^2, 120^2\}$ grid points with a $\text{CFL}=2/\sqrt{2}$.
 \begin{figure}[t!]
	 	\includegraphics[width=\columnwidth]{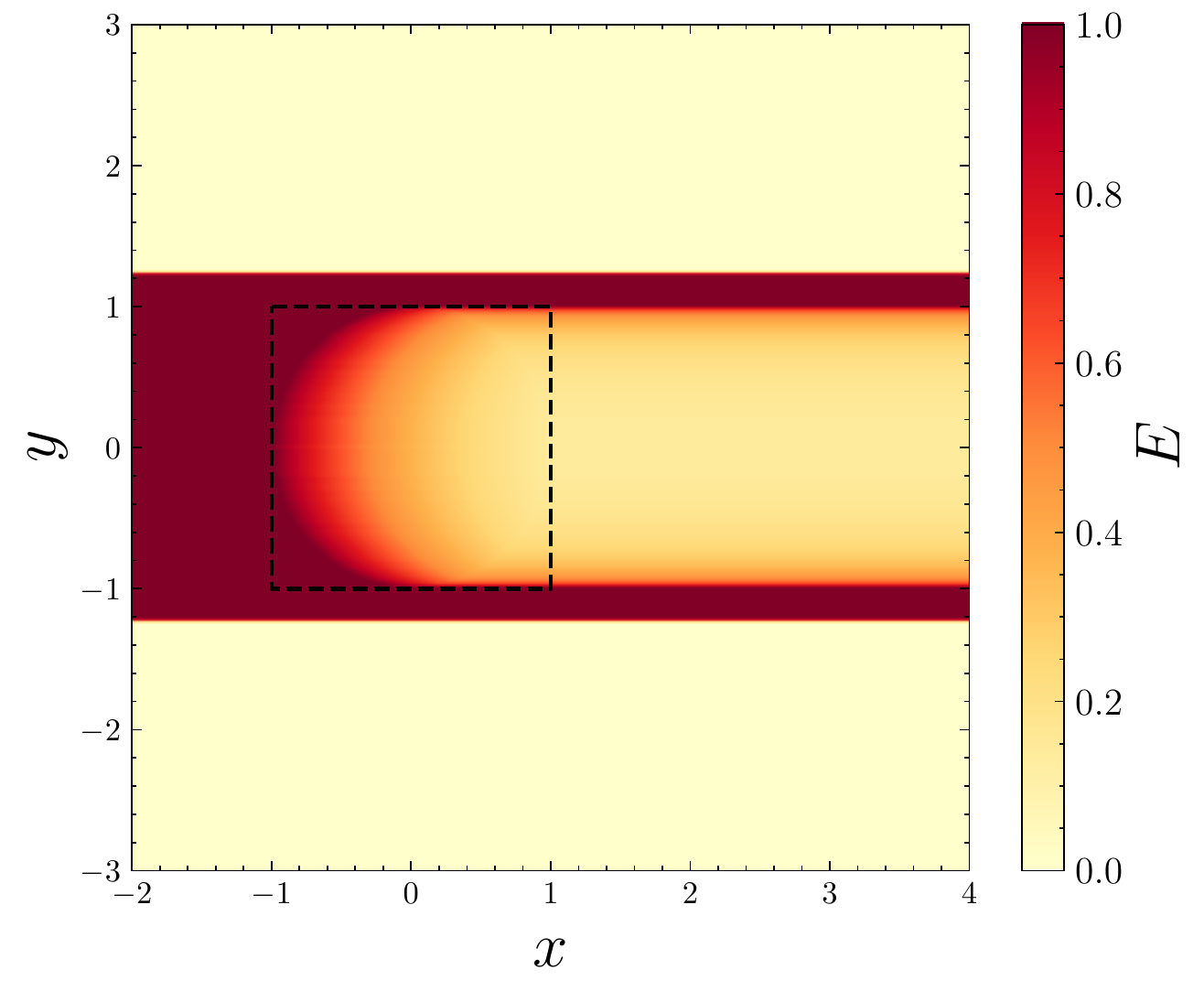}
	 	\caption{{\em Test 4. Shadow Test}. 	
	 		Initially, both the energy density and the radiation fluxes are zero everywhere except in a region on the left boundary. This injection of radiation interacts with a semi-transparent cylinder of radius $R_{\star}=1$ centered at the origin, with an inner opacity of $\kappa_{a}=1$. The energy density is shown at time $t=10$, when the solution has already relaxed to its stationary state. Our numerical scheme perfectly captures the shadow without any lateral diffusion nor oscillations in the radiation field in the vicinity of the cylinder. We have also also placed an extra FMR level (dashed black box), with twice the resolution of the coarsest level, placed at $(x,y) \in [-1,1]^2$, obtaining similar results.}
	 	\label{T4_final}
 \end{figure}

\begin{figure}[t]
	 	\includegraphics[width=\columnwidth]{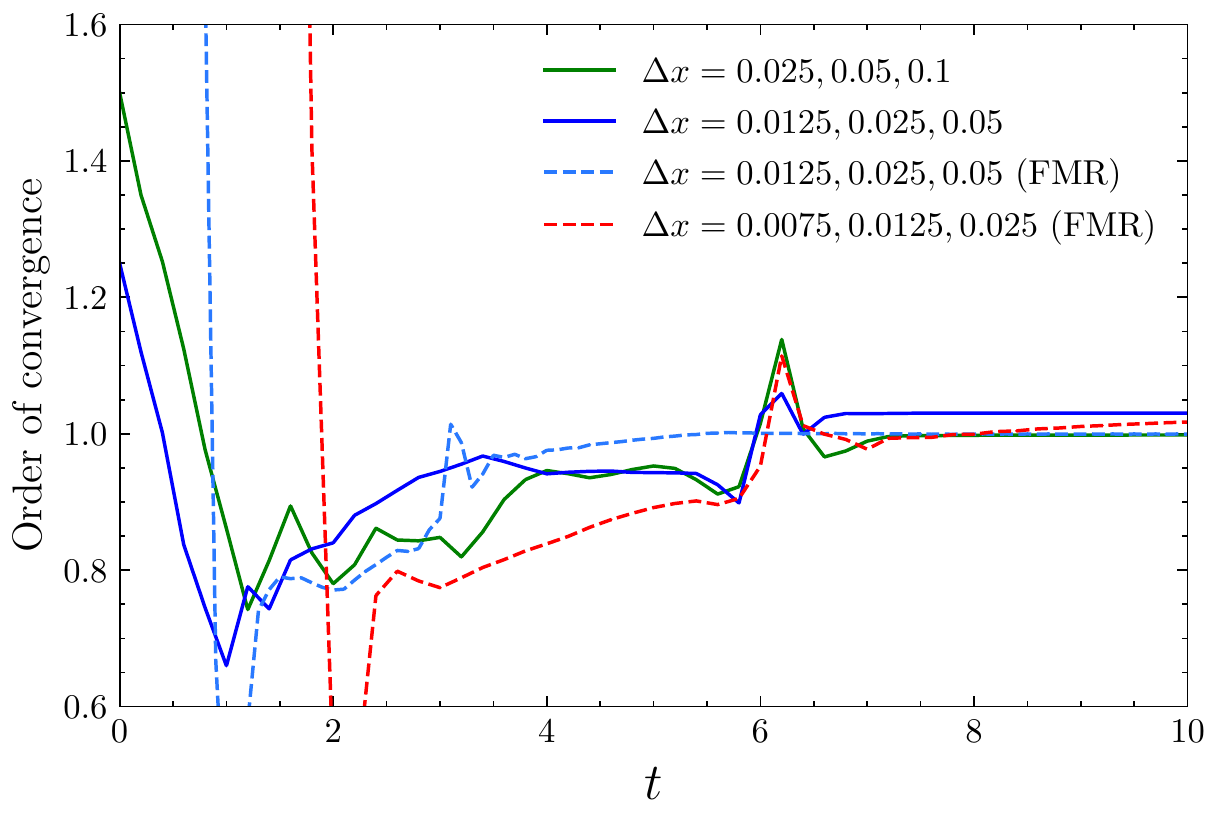}
	 	\caption{{\em Test 4. Shadow Test}. Numerical convergence for several resolutions. We have also considered an additional FMR level, with twice the resolution of the coarsest level, placed around the semi-transparent cylinder, at $ (x,y) \in [-1,1]^2$. 
	 	}
	 	\label{T4_num_conv}
 \end{figure}

 Fig.~\ref{T4_final} shows the radiation energy density for our finest resolution $\Delta x = \Delta y = 0.0125$ at time $t=10$, when the solution has already reached the steady state. We can observe that our numerical scheme is able to perfectly capture the shadow without any lateral diffusion or oscillations in the radiation field in the vicinity of the cylinder. We have also considered the same setup, but with an additional FMR level, with twice the resolution of the coarsest level, placed at $(x,y) \in [-1,1]^2$. As it is displayed  in Fig.~\ref{T4_num_conv}, first order convergence is achieved for all the resolutions presented without and with FMR. This is explained by the highly discontinuous profile of the solution, similar to what we found in Test 1.
 
\subsection*{Test 5. Radiating and Absorbing Sphere}\label{T5}
Finally, we consider a homogeneous radiating and absorbing sphere (see, e.g., \Mycitep{smit97}; \Mycitep{pons00}; \Mycitep{rampp02}; \Mycitep{rad13}; \Mycitep{anninos20}; \Mycitep{weih20}; \Mycitep{chan20}; \Mycitep{rad2022}) first proposed by \Mycite{smit97}, whose analytical solution is known. It reports a scenario that can typically be found in various astrophysical contexts, and that could be a simple model for an isolated, radiating NS. The basic setup is a static (i.e., $v^i(\mathbf{x})=0$), spherically symmetric, homogeneous sphere with constant energy density of radius $R_{\star}=1$, radiating in equilibrium into a surrounding vacuum region.  In this idealized case, we consider that the only interaction process taking place is an isotropic thermal absorption and emission, that we take to be 
\begin{eqnarray}
	&\kappa_{a}(\mathbf{x}, \mathbf{y}, \mathbf{z}) =& \eta(\mathbf{x},\mathbf{y}, \mathbf{z}) =			
	\begin{cases}
		10, &  r \leq R_{\star}\\
		~~0, &  r > R_{\star} 
	\end{cases} \\
	&\kappa_{s}(\mathbf{x},\mathbf{y}, \mathbf{z}) =& \mathbf{0} \,, 
\end{eqnarray}
where $r= \sqrt{x^2 + y^2 + z^2}$.

Under such conditions, the distribution function, $f(r,\mu)$, for this model is known analytically \Mycite{smit97} in terms of the radius $r$ and of the azimuthal angle $\theta$, i.e.,
\begin{equation}
f(r,\mu) = B\left[1-e^{-\kappa_a s(r,\mu)}\right] \,,
\end{equation}
where $\mu \equiv \rm{cos}\,\theta$ and $B$ is a constant that can be freely specified. The function, $s(r,\mu)$, is defined as 
\begin{equation}
s\equiv \begin{cases}
r\mu + R_{\star}g(r,\mu)\,, & \text{if}~ r<R_{\star} ~\quad ~ -1\leq \mu \leq 1, \\
2R_{\star}g(r,\mu)\,, & \text{if}~ r\geq R_{\star} ~\quad ~ \sqrt{1-(R_{\star}/r)^2} \leq
\mu \leq 1 \\
\end{cases}
\end{equation}
and
\begin{equation}
g(r,\mu) \equiv \sqrt{ 1-\left(\frac{r}{R_{\star}}\right)^2(1-\mu^2)} \,.
\end{equation}
By integrating, we obtain the radiation energy density, $J$, measured in the fluid-rest frame as follows 
\begin{equation}
	J(r) = \frac{1}{2} \int_{-1}^1 d\mu ~f(r,\mu) \,.
\label{eq_J_T5}
\end{equation}
Since the fluid velocity is zero, obtaining $E$ is straightforward, leading to the reference exact solution.
\begin{figure}[t!]
	\includegraphics[width=\columnwidth]{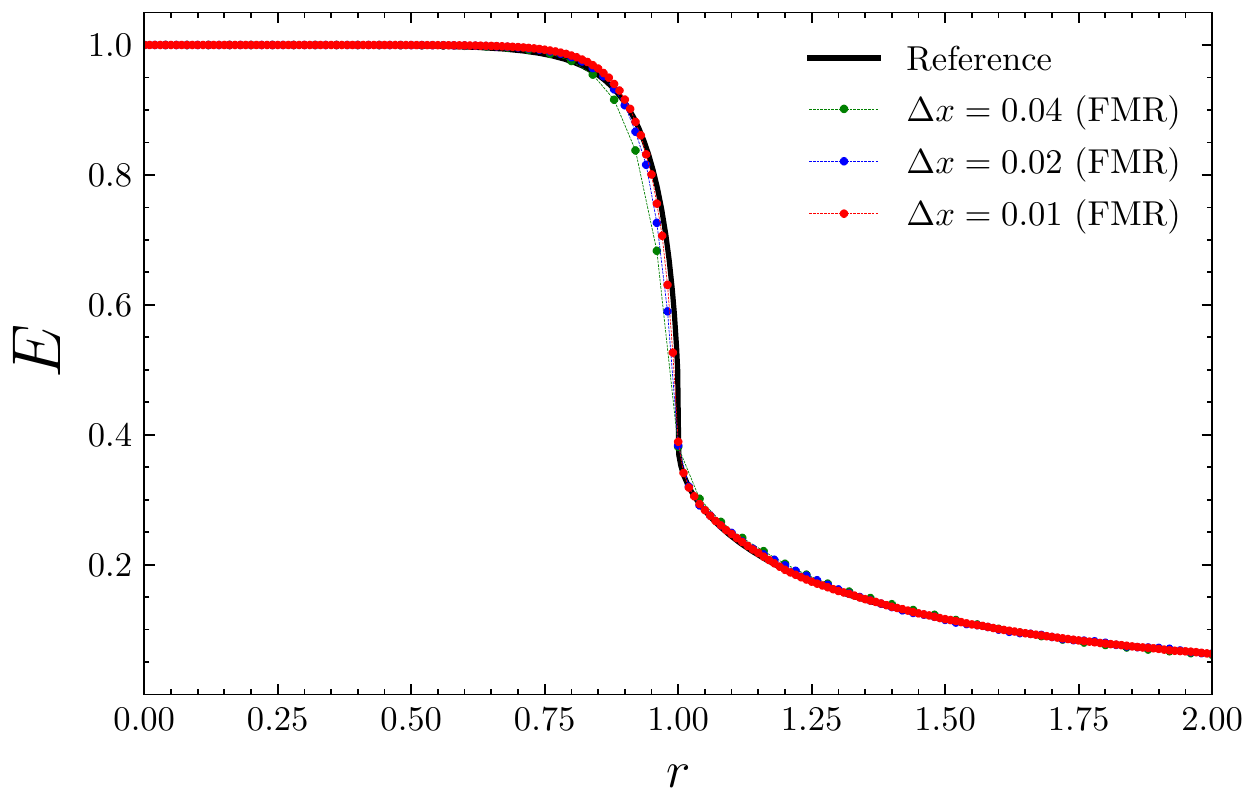}
	\caption{{\em Test 5. Radiating and Absorbing Sphere}. Final profile of the energy density from an homogeneous absorbing ($\kappa_{a}=10$) and emitting ($\eta = 10$) sphere of radius $R_{\star} = 1$. The figure shows the solution at $t=10$ for different resolutions, without and with an extra FMR level, with twice the resolution of the coarsest level, placed at $ (x,y,z) \in [-2,2]^3$.}
	\label{T5_final}
\end{figure}
\begin{figure}[h!]
	\includegraphics[width=\columnwidth]{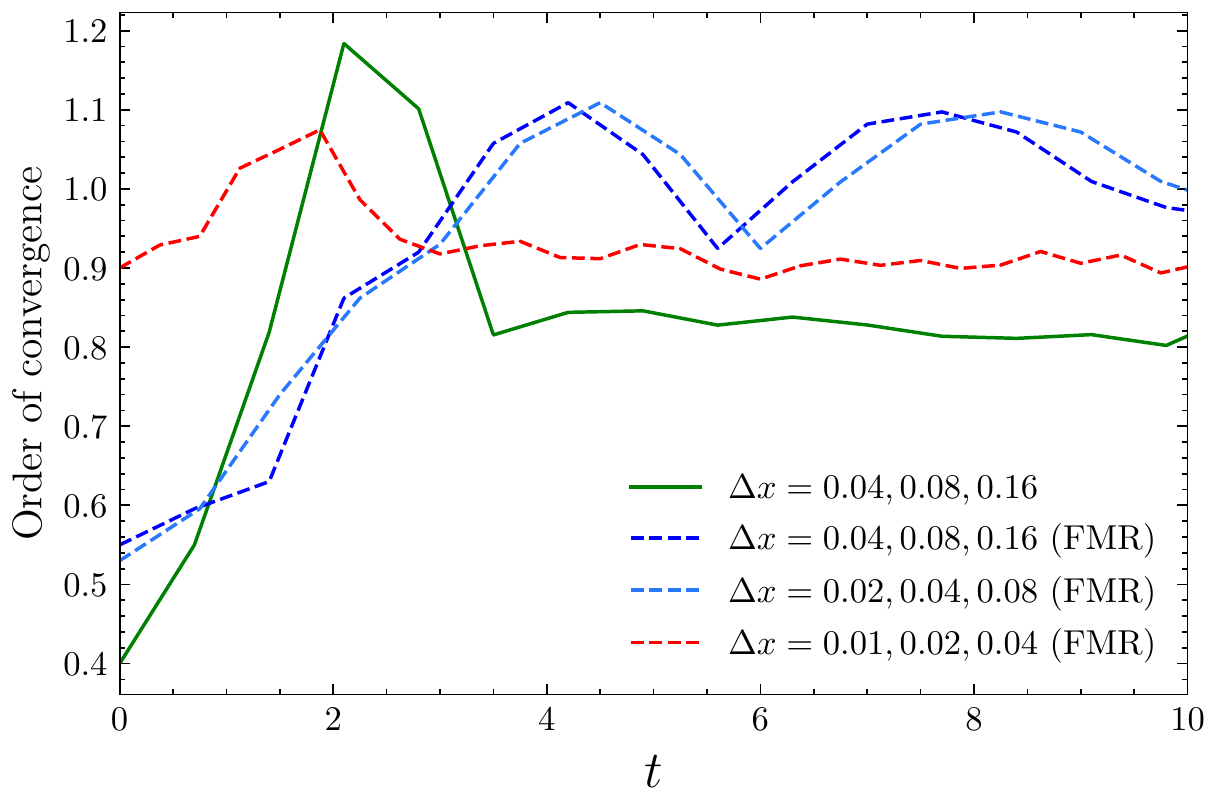}
	\caption{{\em Test 5. Radiating and Absorbing Sphere}. Numerical convergence for several resolutions without and with an extra FMR level, with twice the resolution of the coarsest level, placed at $(x,y,z) \in [-2,2]^3$. First order convergence is achieved due to the discontinuities and shocks produced during the transient to the final solution.}
	\label{T5_num_conv}
\end{figure}

We consider a three-dimensional Cartesian domain with $(x,y,z) \in [-4,4]^3$ and resolved with different resolutions, corresponding to $\{200^3, 100^3, 50^3\}$ grid points with a $\text{CFL}=0.4$.

The radiation energy density for various resolutions at the final time $t=10$, when the solution has already reached the steady state, is shown in Fig. \ref{T5_final}. We have explored again the effect of adding an additional FMR grid, with twice the resolution of the coarsest level, placed at $(x,y,z) \in [-2,2]^3$. In this test, the numerical solution of the M1 approximates quite well the exact reference solution of the full radiative transfer equations, as it was already observed in previous works (see for instance \Mycitep{rad2022}). Finally, we present some convergence tests in Fig.~\ref{T5_num_conv} where we can observe the expected first order convergence due to the highly discontinuous initial setup, both without and with an extra level of refinement.
%
\section{\textbf{Conclusions}}\label{S_conclusions}
We have presented a global high-order numerical scheme for the time evolution of the general relativistic radiation magneto-hydrodynamics equations, which has been implemented in the high-performance code \texttt{MHDuet} \Mycite{mhduet}. Our main goal is to achieve a realistic and accurate model of the neutrino radiation while preserving the high-order numerical schemes commonly employed for the treatment of Einstein and MHD equations. In this work we have implemented the truncated moment formalism (\Mycitep{thorne81}; \Mycitep{shi11}). In particular, we have considered only the first two energy-integrated moments with the Minerbo closure \Mycite{minerbo79}. This hyperbolic-diffusion system with potentially stiff terms arising from the neutrino-matter interaction is commonly known as the M1 scheme. 

The novel part of our work is based on the flexibility of Implicit-Explicit (IMEX) Runge-Kutta schemes to deal with several evolution systems having different accuracy and stability requirements. We have provided a recipe for designing an implicit \emph{tableau} that fits the commonly employed explicit RK schemes. We relax the usual $L$-stability condition required in IMEX methods by also considering $L(\alpha)$-stable schemes for the implicit part, allowing us to achieve higher order schemes with the same number of stages. We would also like to emphasize that one of our requirements for the global time discretization scheme is to efficiently handle the FMR/AMR levels and not to further restrict the CFL condition for the time step. These requirements are important to minimize the cost of computational resources. 

We have managed to find three IMEX schemes that are perfectly suited to our requirements (summarized in Appendix \ref{A_summary_IMEX}). Two of these IMEX schemes contain the explicit \emph{tableau} of the standard $\mathbf{RK4}$ \Mycite{originalRK4}: the third order, $\mathbf{IMEX43}$, and the second order, $\mathbf{IMEX42L}$. Both have a $\text{CFL}$ value close (or equal) to the one corresponding to the explicit \emph{tableau}. In the numerical tests presented here, we have used the $\mathbf{IMEX42L}$ because it is $L$-stable while the $\mathbf{IMEX43}$ is only $L(\alpha)$-stable. However, we have found that this requirement does not introduce any significant difference in the stability of the solutions, except a slight decrease in the CFL condition. We have also found the $\mathbf{IMEX32L}$ which is the unique, second-order and $L$-stable IMEX, satisfying our requirements, that couples with the explicit \emph{tableau} $\mathbf{SSP-RK3}$ \Mycite{shu_osher}. In addition, we have provided a recipe for the design of these IMEX schemes, as well as explained in detail the implicit treatment of stiff source terms and the numerical methods employed for modeling the neutrino radiation.

We have performed several numerical tests to evaluate our implementation of the M1 scheme. We follow the recent work of \Mycite{rad2022} as it encompasses in a set of tests the main physical aspects appearing in the post-merger of BNS merger simulations. The results confirm the robustness of our numerical scheme implemented in \texttt{MHDuet}. We achieve the expected second order convergence for smooth solutions.
Future improvements include the development of more efficient schemes which avoid iterative solvers following the semi-implicit ideas in \Mycite{boscarino:hal-00983924}.  

The described implementation is the starting point for detailed studies, through numerical simulations, of BNS mergers, including not only a realistic description of neutrino radiation, but also large-eddy-simulations (LES) techniques to accurately resolve the magneto-hydrodynamic instabilities that develop during these mergers. 

\appendix	
\section{Summary of IMEX schemes}\label{A_summary_IMEX}
We report in this appendix the results of the Butchers \emph{tableau}  of the IMEX schemes that satisfy our requirements, such that the explicit RK is one of the commonly used in Numerical Relativity applications, while the implicit RK is at least second order, $L(\alpha)$-stable and stiffly accurate. The double Butcher \emph{tableau} for the \textbf{IMEX43(4,4,3)}, \textbf{IMEX42L(4,4,2)} and \textbf{IMEX32L(3,3,2)} are displayed respectively in Tables \ref{A_final_IMEX43}, \ref{A_final_IMEX4} and \ref{A_final_IMEX3}.
\begin{table}[t!]
	\caption{\emph{Tableau} for the explicit (left) and implicit (right) \textbf{IMEX43(4,4,3)} scheme \label{A_final_IMEX43}}
	\begin{minipage}{1.1in}
		\begin{tabular} {c c c c c c}
			0   & \vline  &  0  &  0  & 0 & 0 \\
			1/2   & \vline  &  1/2  &  0  & 0 & 0 \\
			1/2 & \vline   & 0 & 1/2 & 0 & 0 \\
			1  & \vline   & 0 & 0 & 1 & 0 \\
			\hline 
			& \vline &  1/6 & 1/3 & 1/3 & 1/6 \\
		\end{tabular}
	\end{minipage}~~~~~~~~~
	\begin{minipage}{1.1in}
		\begin{tabular} {c |c c c c c}
			$0$ &  $0$ & $0$ & $0$ & $0$\\
			$1/2$ & $1/3$ & $1/6$ & $0$ & $0$ \\  
			$1/2$ & $1/6$& $4/21$ &$1/7$& $0$\\  
			$1$ & $1/6$ & $1/3$ & $1/3$ & $1/6$ \\  
			\hline \\[-1.0em] 
			& $1/6$& $1/3$ & $1/3$ & $1/6$ \\
		\end{tabular}
	\end{minipage}
\end{table}
\begin{table}[t!]
	\caption{\emph{Tableau} for the explicit (left) and implicit (right) \textbf{IMEX42L(4,4,2)} scheme \label{A_final_IMEX4}}
	\begin{minipage}{1.1in}
		\begin{tabular} {c c c c c c}
			0   & \vline  &  0  &  0  & 0 & 0 \\
			1/2   & \vline  &  1/2  &  0  & 0 & 0 \\
			1/2 & \vline   & 0 & 1/2 & 0 & 0 \\
			1  & \vline   & 0 & 0 & 1 & 0 \\
			\hline 
			& \vline &  1/6 & 1/3 & 1/3 & 1/6 \\
		\end{tabular}
	\end{minipage}~~~~~~~~~
	\begin{minipage}{1.1in}
		\begin{tabular} {c |c c c c c}
			$0$ &  $0$ & $0$ & $0$ & $0$\\
			$1/2$ & $1/4$ & $1/4$ & $0$ & $0$ \\  
			$1/2$ & $0$& $1/6$ &$1/3$& $0$\\  
			$1$ & $1/6$ & $1/3$ & $1/3$ & $1/6$ \\  
			\hline \\[-1.0em] 
			& $1/6$& $1/3$ & $1/3$ & $1/6$ \\
		\end{tabular}
	\end{minipage}
\end{table}
\begin{table}[t!]
	\caption{\emph{Tableau} for the explicit (left) and implicit (right) \textbf{IMEX32L(3,3,2)} scheme \label{A_final_IMEX3}}
	\begin{minipage}{1.1in}
		\begin{tabular} {c | c c c}
			$0$ &  $0$ & $0$ & $0$ \\
			$1$ & $1$ & $0$ & $0$ \\  
			$1/2$ & $1/4$ & $1/4$ & $0$ \\  
			\hline \\[-1.0em] 
			& $1/6$& $1/6$ & $2/3$ \\
		\end{tabular}
	\end{minipage}~~~~~~~~~
	\begin{minipage}{1.1in}
		\begin{tabular} {c |c c c}
			$0$ &  $0$ & $0$ & $0$ \\
			$1$ & $1/2$ & $1/2$ & $0$ \\  
			$1/2$ & $1/6$ & $1/6$ & $2/3$ \\  
			\hline \\[-1.0em] 
			& $1/6$& $1/6$ & $2/3$ \\
		\end{tabular}
	\end{minipage}
\end{table}

The stability region for the three different implicit and the two explicit schemes discussed here are displayed in Fig.~\ref{stability_functions}. In general, the stability region of the explicit schemes is the interior of the closed curve, while that of the implicit schemes is the exterior. We confirm that both \textbf{I42L(4,2)} and \textbf{I32L(3,2)} are $L$-stable, and that \textbf{I43(4,3)} is only $L(\alpha)$-stable. For the latter there is a slight crossover with the stability region of \textbf{RK4(4,4)}, which is responsible for the slight decrease in the maximum allowed CFL factor.

We have performed several tests with a simple advection-diffusion equation to check the largest allowed value of the CFL leading to stable simulations, confirming that these three IMEX are the ones that meet all our requirements.
\section{Comparison between different time discretization schemes}\label{appendix_comparison_sec}
 \begin{figure*}[t!]
	\centering
	\subfigure{\includegraphics[scale=1.0]{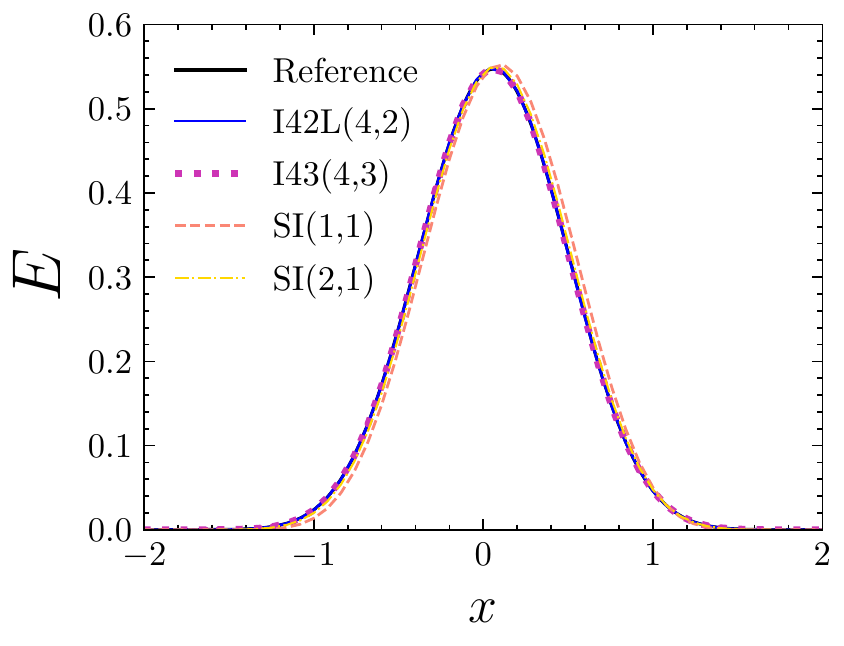}}\quad
	\subfigure{\includegraphics[scale=1.0]{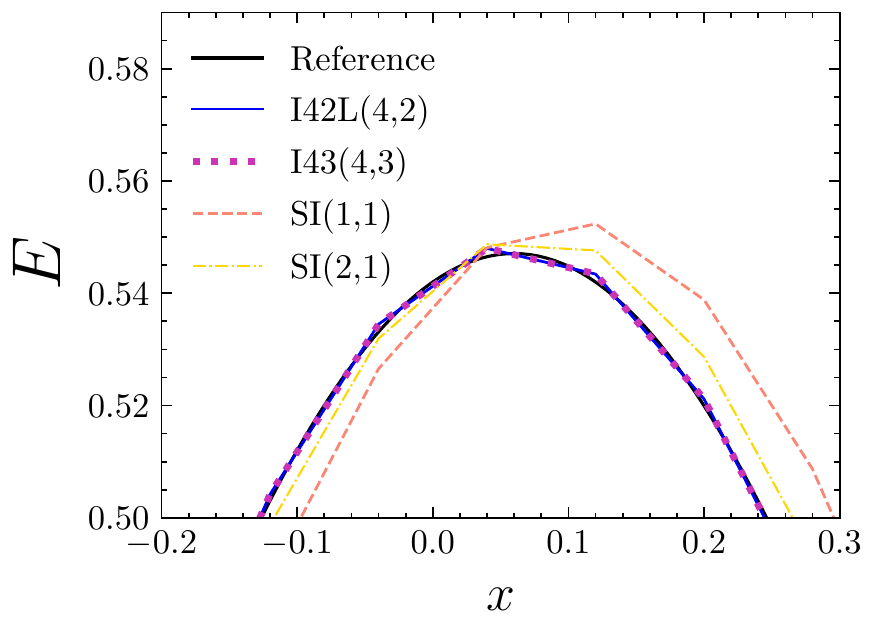}}
	\caption{{\em Appendix \ref{appendix_comparison_sec} - Comparison of IMEX vs semi-implicit schemes. } The energy density is initialized as a Gaussian pulse of radiation which is advected while promptly diffuses in an intermediate scattering media with $\kappa_{s} = 10$. We are using fourth order accurate central finite differences (CD4), although with second order show similar (but slightly smaller) differences. The numerical solution is presented for different time integration schemes. 	\label{appendix_figure}}
\end{figure*}
In this appendix we have compared the accuracy of the IMEX schemes presented in this work (see Appendix \ref{A_summary_IMEX}) and other semi-implicit schemes widely employed in the literature. We have noticed that the tests performed in Sec. \ref{S_numerical_tests} are not appropriate for a fair comparison between different temporal integration schemes, since the initial data is often discontinuous and in many situations the order of convergence is constrained by the spatial discretization scheme chosen in Sec. \ref{spatial_discretization_scheme}, which is second order accurate in the best case. The semi-implicit schemes we will consider are the \textbf{SI(1,1)} and the \textbf{SI(2,1)}. Their Butcher tables in CK form and enforcing that they are stiffly accurate (i.e., $w_i=a_{si}$, $i=1,\ldots,s$) are presented in Table \ref{SI(1,1)} and \ref{SI(2,1)} respectively.
\begin{table}[t!]
	\caption{\emph{Tableau} for the explicit (left) and implicit (right) \textbf{SI(1,1)} scheme \label{SI(1,1)}}
	\begin{minipage}{1.1in}
		\begin{tabular} {c c c c}
			0 & \vline   & 0 & 0 \\
			1  & \vline   & 1 & 0 \\
			\hline 
			& \vline &  1 & 0 \\
		\end{tabular}
	\end{minipage}~~~~~~~~~
	\begin{minipage}{1.1in}
		\begin{tabular} {c |c c c}
			$0$ & $0$& $0$\\  
			$1$ & $0$ & $1$ \\  
			\hline \\[-1.0em] 
			& $0$& $1$ \\
		\end{tabular}
	\end{minipage}
\end{table}
\begin{table}[t!]
	\caption{\emph{Tableau} for the explicit (left) and implicit (right) \textbf{SI(2,1)} scheme \label{SI(2,1)}}
	\begin{minipage}{1.1in}
		\begin{tabular} {c c c c c}
			0   & \vline   &  0  & 0 & 0 \\
			1/2 & \vline   & 1/2 & 0 & 0 \\
			1  & \vline   & 0 & 1 & 0 \\
			\hline 
			& \vline &  0 & 1 & 0 \\
		\end{tabular}
	\end{minipage}~~~~~~~~~
	\begin{minipage}{1.1in}
		\begin{tabular} {c |c c c c}
			$0$ & $0$ & $0$ & $0$ \\  
			$1/2$ & $0$ &$1/2$& $0$\\  
			$1$ & $0$ & $0$ & $1$ \\  
			\hline \\[-1.0em] 
			& $0$&  $0$ & $1$ \\
		\end{tabular}
	\end{minipage}
\end{table}
The semi-implicit scheme \textbf{SI(1,1)} (see, e.g., \Mycitep{sado13}; \Mycitep{fou15}; \Mycitep{oconnor15};  \Mycitep{kuroda16}; \Mycitep{foucart_review}) is first order for both implicit and explicit terms. Another popular semi-implicit scheme for transport problems is \textbf{SI(2,1)} (see, e.g., \Mycitep{mcclarren08}; \Mycitep{rad13}; \Mycitep{just15}; \Mycitep{rad2022}) which is second order accurate for explicit terms and first order for implicit terms. 

For this comparison, we have decided to simplify the evolution equations (eqs. \ref{eq_evolE}, \ref{eq_evolF}) by assuming that the spacetime is flat, the fluid is at rest and the problem is one-dimensional. The resulting set of equations is as follows,
\begin{eqnarray}
	&\partial_t E  \label{reduced_evol}
	+ \partial_x F^x &= (\eta - \kappa_a E) ~,  \\
	&\partial_t F^x  \nonumber
	+\partial_x {P^{x}}_{x} &= -(\kappa_a + \kappa_{s})F^{x}.  
\end{eqnarray}
Here we will consider only diffusion (i.e., $\eta(\mathbf{x}) = \kappa_a(\mathbf{x}) = 0$). We initialize the system considering the propagation of a gaussian profile, namely
\begin{align}
	&E(t=0, ~\mathbf{x}) = e^{-9x^2} \,, \\
	&F^x(t=0, \mathbf{x}) = \frac{1}{\sqrt{3}}E(t=0, ~\mathbf{x}) \,.
\end{align}
The simulation is performed in a Cartesian domain with $x\in [-3,3]$ and solved by different resolutions, corresponding to \{300,150,75\} grid points with a $\text{CFL}=0.5$. 
%
%
%
\begin{table*}
	\begin{ruledtabular}
		\begin{tabular}{c|c||ccccc}
			\centering
			\multirow{2}{*}{\textbf{Spatial scheme}} & \multirow{2}{*}{\textbf{Scattering coefficient ($\kappa_{s}$)}} & \multicolumn{4}{c}{\textbf{Numerical order of convergence}} & 
			\\ 
			 &  & {\textbf{I42L(4,2)}} & \textbf{I43(4,3)} &
			 \textbf{SI(1,1)} & \textbf{SI(2,1)} &
			\\ \hline\hline 			
			\multirow{4}{*}{\emph{CD2}} & 0     & 2   &  2   & 1.5   & 2  & \\ 
			 & 10 & 2   & 2   & 1 & 1  & \\ 
			 & 100 & 2   & 2   & 1 & 1  &  \\ 
			 & 1000 & 2   & 2   & 1 & 1.5  &   \\ 
			 \hline\hline 			
			\multirow{4}{*}{\emph{CD4}} & 0     & 4   &  4   & 1.3   & 2  & \\ 
			& 10 & 4   & 4   & 1 & 1.3  & \\ 
			& 100 & 2   & 3   & 1 & 1  &  \\ 
			& 1000 & 3.5   & 4   & 1 & 1  &   
		\end{tabular}
		\caption{{\em Numerical order of convergence} for different values of the scattering coefficient $\kappa_{s}=(0,10,100,1000)$ using second and fourth order central finite differences (CD2 and CD4) with different time integration schemes.	}\label{convergence_appendix}
	\end{ruledtabular}
\end{table*}

We have changed the spatial discretization scheme presented in Sec. \ref{spatial_discretization_scheme} for second and fourth order accurate central finite differences (CD2 and CD4) with their corresponding Kreiss-Oliger dissipation operators (\Mycitep{gustafsson_book}). The numerical orders of convergence obtained at time $t=2$ are presented in Table \ref{convergence_appendix} for different values of the scattering coefficient,
\[
	\kappa_{s}(\mathbf{x}) = [0, 10, 100, 1000] ~.
\]
This test allows us to validate the expected order of convergence when no implicit source terms are present (i.e., for $\kappa_{s}(\mathbf{x}) = 0$), as well as for intermediate and high scattering regimes.
In Fig. \ref{appendix_figure} we show the final profile at time $t=2$ obtained with our lowest resolution ($\Delta x = 0.08$) with CD4 for an intermediate scattering regime ($\kappa_{s} = 10$). It becomes pretty clear from this figure and the results of Table \ref{convergence_appendix} that the \textbf{I42L(4,2)} and \textbf{I43(4,3)} behave quite well in all regimes. The semi-implicit schemes \textbf{SI(1,1)} and \textbf{SI(2,1)} suffer due to: (i) the first order accuracy in regimes where implicit terms dominate, and (ii) the CFL stringent constraint. Finally, we would like to recall that the CD4 spatial discretization scheme could not be used in this problem, since we require the fluxes to be reduced to a second-order finite difference scheme in the optical thick limit. However, the use of higher order IMEX schemes assures us that the accuracy of the neutrino equations will only be limited by the accuracy of the spatial discretization scheme.
%
\section{Jacobian full expression}\label{A_jacobian_terms}
\emph{\textbf{Note:} Full expression of the Jacobian as published in \Mycite{rad2022}.}
Let us define $\kappa_{as} = \kappa_a + \kappa_s$
and write $J,H_i$ as a function of $E,J_i,\chi$ for the Minerbo closure, namely
{ \small
\begin{eqnarray}\label{Jminerbo}
&& J (E,F_i) = B_0 + d_{\text{thin}} B_{\text{thin}} + d_{\text{thick}} B_{\text{thick}} \\
\label{Hminerbo}
&&H_i (E,F_i) = - (a_{v 0} + d_{\text{thin}} a_{v \text{thin}} + d_{\text{thick}} a_{v \text{thick}} ) v_ i  \nonumber \\
&&- d_{\text{thin}} a_{f \text{thin}} {\hat f}_i - (a_{F 0} + d_{\text{thick}} a_{F \text{thick}} ) F_i 
\end{eqnarray}
}
where we have also defined
{ \small
\begin{eqnarray} 
&& {\hat f}_i = \frac{F_i}{\sqrt{F_k F^k}} ~,~
d_{\text{thick}} = \frac{3}{2} (1- \chi) ~,~
d_{\text{thin}} = 1 - d_{\text{thick}}~,  \\
&& B_0 = W^2 \left[ E - 2 v_k F^k \right] ~~,\\
&& B_{\text{thin}} = W^2 E (v_k {\hat f}^k)^2 ~~,\\
&& B_{\text{thick}} = \frac{W^2-1}{2 W^2 + 1} \left[
4 W^2 (v_k F^k) + (3 - 2 W^2) E  \right] ~~,\\
&& a_{v 0} = W B_0 ~~,\\
&& a_{v \text{thin}} = W B_{\text{thin}} ~~,\\
&& a_{v \text{thick}} = W B_{\text{thick}} \\
&& + \frac{W}{2 W^2 + 1}
\left[ (2 W^2 - 1) (v_k F^k) + (3 - 2 W^2) E \right] ~~, \nonumber \\
&& a_{f \text{thin}} = W E (v_k {\hat f}^k) ~~, \\
&& a_{F 0} = - W ~~,\\
&& a_{F \text{thick}} =  W v^2  ~~.
\end{eqnarray}
}
The Jacobian ${\cal J} \equiv \left(\frac{\partial { R_V}  }{\partial {\bf V}}\right)$ of the undensitized fields is then given by
{\small
\begin{eqnarray}
{\cal J}_{00} &=& -\alpha W \left( \kappa_{as} - \kappa_{s} 
\frac{\partial J}{\partial E} \right) ~,\\
{\cal J}_{0j} &=& +\alpha W \left( \kappa_{s} \frac{\partial J}{\partial F_j} + \kappa_{as} v^j 
\right) ~,\\
{\cal J}_{i0} &=& -\alpha \left( \kappa_{as} \frac{\partial H_i}{\partial E}
+ W \kappa_{a} \frac{\partial J}{\partial E} v_i
\right) ~,\\
{\cal J}_{ij} &=& -\alpha \left( \kappa_{as} \frac{\partial H_i}{\partial F_j}
+ W \kappa_{a} v_i \frac{\partial J}{\partial F_j} 
\right) ~.
\end{eqnarray}
}
where the necessary derivatives are
\begin{widetext}
	{\small
	\begin{align}
	\scriptsize	
	&&\frac{\partial J}{\partial E} = W^2 + d_{\text{thin}} 
	(v_k {\hat f}^k)^2 W^2 
	+ d_{\text{thick}} \frac{(3-2 W^2)(W^2 - 1)}{1 + 2 W^2}~,\\
	&&\frac{\partial J}{\partial F_j} = 2 W^2 \left( 
	-1 + d_{\text{thin}} \frac{E (v_k {\hat f}^k)}{F} 
	+ 2 d_{\text{thick}} \frac{W^2 - 1}{1 + 2 W^2}  \right) v^j
	- 2 d_{\text{thin}} \frac{W^2 E (v_k {\hat f}^k)^2}{F}
	{\hat f}^j  ~,\\
	&&\frac{\partial H_i}{\partial E} = W^3 \left( 
	-1 - d_{\text{thin}} (v_k {\hat f}^k)^2 
	+ d_{\text{thick}} \frac{2 W^2 - 3}{1 + 2 W^2}  \right) v_i
	- d_{\text{thin}} W (v_k {\hat f}^k) {\hat f}_i  ~,\\
	&&\frac{\partial H_i}{\partial F_j} = W \left( 
	1 - d_{\text{thin}} \frac{E (v_k {\hat f}^k)}{F} 
	- d_{\text{thick}} v^2  \right) \delta^j_i
	+ 2 W^3 \left[ 1 - d_{thin} \frac{E (v_k {\hat f}^k)}{F} 
	- d_{\text{thick}} \left( v^2 + \frac{1}{2 W^2 (1 + 2 W^2)}  \right)     \right] v_i v^j \nonumber \\
	&&+ 2 d_{\text{thin}}  \frac{W E (v_k {\hat f}^k)}{F}  {\hat f}_i {\hat f}^j
	+ 2 d_{\text{thin}}  \frac{W^3 E (v_k {\hat f}^k)^2}{F}  v_i {\hat f}^j
	- d_{\text{thin}}  \frac{W E}{F}  {\hat f}_i v^j  ~~.
	\end{align}
}%
\end{widetext}
%
\section*{\textbf{Acknowledgments}}
We are indebted with David Radice for many useful discussions on the neutrino transport problem. We also thank Martin Obergaulinger for many clarifications on the subject. Finally, it is a pleasure to thank Miguel Bezares for reviewing an earlier version of the manuscript. CP acknowledges the hospitality at the Institute for Pure \& Applied Mathematics (IPAM) through the Long Program ``Mathematical and Computational Challenges in the Era of Gravitational Wave Astronomy", where this project was initiated.
MRI thanks financial support PRE2020-094166 by MCIN/AEI/PID2019-110301GB-I00 and by ``FSE invierte en tu futuro''.
LP would like to thank the Italian Ministry of Instruction, University and
Research (MIUR) to support this research with funds coming from PRIN
Project 2017 (No. 2017KKJP4X entitled “Innovative numerical methods for
evolutionary partial differential equations and applications”).
This work was supported by the Grant PID2019-110301GB-I00 funded by MCIN/AEI/10.13039/501100011033 and by "ERDF A way of making Europe" (JM and CP).
%
\section*{\textbf{Data Availability}}
Data generated for this study will be made available upon reasonable
request to the corresponding authors.
%
\bibliographystyle{apalike}
\bibliography{refs}
\end{document}